\documentclass[twocolumn,times,tighten]{aastex631}
\usepackage{newtxtext,newtxmath}

\usepackage[T1]{fontenc}
\usepackage{ae,aecompl}
\usepackage{braket}
\usepackage{multirow}

\usepackage{etoolbox}
\makeatletter
\patchcmd\@combinedblfloats{\box\@outputbox}{\unvbox\@outputbox}{}{

}
\makeatother

\usepackage{graphicx}	
\usepackage{amsmath}	
\usepackage{amssymb}	
\usepackage{float}
\usepackage{verbatim}

\newcommand{\alpro}{\textsc{ALPro}}

\newcommand{\g}{g_{a\gamma}}

\definecolor{C0}{HTML}{1f77b4}
\definecolor{C1}{HTML}{ff7f0e}
\definecolor{C2}{HTML}{2ca02c}
\definecolor{C3}{HTML}{d62728}
\definecolor{C6}{HTML}{e377c2}

\shorttitle{The role of calibration in \textit{Athena}'s projected bounds on axions}
\shortauthors{J.~Sisk-Reyn\'es et al.}

\begin{document}
\title
{
  Physics Beyond the Standard Model with Future X-ray Observatories: \newline \begin{centering}Projected Constraints on Very-Light Axion-Like Particles with \textit{Athena} and \textit{AXIS}
  \end{centering}
}


\newcommand{\ioa}{Institute of Astronomy, University of Cambridge, Madingley Road, Cambridge, CB3 0HA;
\href{mailto:jms332@cam.ac.uk}{jms332@cam.ac.uk}}

\author[0000-0003-3814-6796]{Júlia Sisk-Reynés}
\affiliation{\ioa}

\author[0000-0002-1510-4860]{Christopher S.~Reynolds }
\affiliation{\ioa}
\affiliation{Dept.~of Astronomy, University of Maryland, College Park, MD 20742, USA}

\author[0000-0002-8466-7317
]{Michael~L.~Parker}
\affiliation{\ioa}

\author[0000-0002-3493-7737]{James~H.~Matthews}
\affiliation{Department of Physics, Astrophysics, University of Oxford, Denys Wilkinson Building, Keble Road, Oxford OX1 3RH, UK}
\affiliation{\ioa}

\author[0000-0001-7271-4115]{M.~C.~David Marsh}
\affiliation{The Oskar Klein Centre, Department of Physics, Stockholm University, Stockholm 106 91, Sweden}

\begin{abstract}
\noindent
Axion-Like Particles (ALPs) are well-motivated extensions of the Standard Model of Particle Physics and a generic prediction of some string theories.~X-ray observations of bright Active Galactic Nuclei (AGN) hosted by rich clusters of galaxies are excellent probes of very-light ALPs, with masses $\mathrm{log}(m_\mathrm{a}/\mathrm{eV}) < -12.0$.~We evaluate the potential of future X-ray observatories, particularly \textit{Athena} and the proposed \textit{AXIS}, to constrain ALPs via observations of cluster-hosted AGN, taking NGC~1275 in the Perseus cluster as our exemplar.~Assuming perfect knowledge of instrument calibration, we show that a modest exposure ($200$-$\mathrm{ks}$) of NGC~1275 by {\it Athena} permits us to exclude all photon-ALP couplings $g_\mathrm{a\gamma} > 6.3 \times 10^{-14} \ {\mathrm{GeV}}^{-1}$ at the 95\%~level, as previously shown by \citet[][]{Conlon_2017_ProjectedBoundsOnAthena}, representing a factor of 10 improvement over current limits.~We then proceed to assess the impact of realistic calibration uncertainties on the {\it Athena} projection by applying a standard $Cash$ likelihood procedure, showing the projected constraints on $g_
\mathrm{a\gamma}$ weaken by a factor of 10 (back to the current most sensitive constraints).~However, we show how the use of a deep neural network can disentangle the energy-dependent features induced by instrumental miscalibration and those induced by photon-ALP mixing, allowing us to recover most of the sensitivity to the ALP physics.~In our explicit demonstration, the machine learning applied allows us to exclude $g_\mathrm{a\gamma} > 2.0 \times 10^{-13} \ {\mathrm{GeV}}^{-1}$, complementing the projected constraints of next-generation ALP dark matter birefringent cavity searches for very-light ALPs.~Finally, we show that a 200-ks~\textit{AXIS}/on-axis observation of NGC~1275 will tighten the current best constraints on very-light ALPs by a factor of 3.
\end{abstract}

\section{Introduction}
\label{section:s1_into}
Astrophysical observations probe physical extremes that provide a window onto physics beyond the Standard Model (SM).~Axions and axion-like particles in particular arise from well-motivated extensions to the SM and are amenable to astrophysical study.~The Quantum Chromodynamics (QCD) axion is the leading solution to the so-called ``strong CP problem'' in the SM.~This problem arises due to the unexpectedly small value of the electric dipole moment for the neutron, $< 10^{-9} \ e\mathrm{cm}$ \citep[][]{PhysRev.184.1660_ElectricDipoleNeutronMoment_Predictions1969}, yielding experimental evidence for the conservation of Charge-Conjugation Parity (CP) symmetry by the strong force, which is not predicted by SM theory.~Such a tension can be solved through the introduction of a new $U(1)$ symmetry in the SM Lagrangian which enables replacing the CP-violating phase by a pseudo-scalar, dynamical field $a$ \citep[][]{PecceiQuinnOriginal}.~This symmetry, known as the Peccei-Quinn symmetry, is both spontaneously broken at a high scale and anomalous, resulting in a pseudo-Nambu-Goldstone boson (pNGB) of mass $m_\mathrm{a}$; the QCD axion \citep[][]{SWeinberg_QCDaxion, FWilczek_QCDaxion}.~The dynamics and interactions of such particle can be derived from its mass $m_\mathrm{a}$ and decay constant, whose product is set by that of the mass and decay constant of the neutral pion.~

ALPs are pNGBs that arise in many beyond SM (BSM) frameworks such as string compactifications \citep[][]{greenSchwarzWitten_2012,Conlon_2006_iib, Svrcek_witten_2006_iib, Cicoli_2012_typeiib}.~Excitingly, some of these theories predict sizeable ALP-photon couplings that are within reach of astrophysical observations \citep[][]{Halverson19StringTheories,djemarsh_calabiYau_stringTheory_superradiance, demirtas_20_axiverse}.~ALPs do not couple to QCD, but can couple to other vector bosons and have derivative interactions with matter fields.~Another driver for the rising interest in ALP and axion searches is their potential to be Dark Matter (DM) \citep[][]{PRESKILL_Wise_Wilczek_CosmoInvisibleAxion_83, Abbott_Sikivie_CosmologicalBoundOnAxion_83, Dine_Fischler_83_HarmlessAxion, Irastorza_dm_axion_review}, or, for sufficiently light ALPs, dark energy \citep[][]{ALPsQuintessentialDE_Carroll}.~In the low-energy effective theory relevant for observations, the ALP mass $m_\mathrm{a}$ and ALP coupling to the electromagnetic field $g_\mathrm{a\gamma}$ are assumed to be independent, regardless of their cosmological role.

Most relevant to our work is the study of the interaction of ALPs with electromagnetism.~At a fundamental level, this interaction is described by the Lagrangian term
\begin{equation}
  \centering
\label{equation:eq1_interactionLangrangian_alps}
  \mathcal{L}_\mathrm{a\gamma} = a \, g_\mathrm{a\gamma} \, \textbf{B} \cdot \textbf{E},
\end{equation} where $a$ is the ALP field; $\textbf{B}$ and $\textbf{E}$ are the magnetic and electric fields, respectively; and $g_\mathrm{a\gamma}$ is the photon-ALP coupling constant.~In most astrophysical searches for ALPs, the relevant electric field is that of the photon beam, whereas the magnetic field is imposed externally as a ``background'' field.~The resulting expressions for photon-ALP mixing in the simple case of a uniform magnetic field are outlined in Appendix \ref{section:a1_theory_photonALPmixing}.~Importantly, photon-ALP mixing requires the external \textbf{B}-field component perpendicular to the beam propagation direction to be non-zero.~

Photon-ALP mixing will induce energy-dependent features in the spectrum of a photon beam propagating through an ionised, magnetised plasma as the photons mix with sufficiently light ALPs.~Photon-ALP mixing is an oscillatory function with energy that is strongly dependent on the properties of the plasma, as well as $m_\mathrm{a}, g_\mathrm{a\gamma}$.~It is the presence --or absence-- of such energy-dependent modulations in the spectra of astrophysical sources that allows us to constrain ALPs, that is, to set an upper bound on $g_\mathrm{a\gamma}$ for a given $m_\mathrm{a}$.~

Bounds on anomalous stellar cooling provided the earliest astrophysical bounds on ALPs of $m_\mathrm{a} < 10^{-10} \ \mathrm{eV}$ \citep[][]{Raffelt:1987_MixingOfPhoton, Raffelt:1996_StarsAsLabsForAxions, PierlucaCarenza_HeliumBurningStars_MostRecent}.~Competitive constraints on $g_\mathrm{a\gamma}$ for such ALP masses come from gamma-ray observations of SN1987A \citep[][]{SN1987Bounds} and other core-collapse supernovae \citep[][]{Meyer_2020_FermiLAT_FielfOfView_ALPs_sneExtraGalactic, meyer2021, milena_fermilat_lowenergy_ccsne}.~These limits have been complemented in a broader mass range ($m_\mathrm{a} < 0.02 \ \mathrm{eV}$) by the CERN Axion Solar Telescope \citep[CAST; ][]{CAST_AbsenceOfALPs}, which aimed to reprocess ALPs generated at the core of the Sun back into X-ray photons within a magnetised chamber located at CERN and excluded all photon-ALP couplings $g_\mathrm{a\gamma} > 6.6\times 10^{-11} \ {\mathrm{GeV}}^{-1}$ at the $95\%$ level.~We refer to \citet[][]{IAXO19_ProjectedSensitivity} and \citet[][]{ssaxi_jaesub_cfa} for projected constraints from solar axions from future ground-based and spaced-based next-generation missions.~

Many astrophysical settings are both photon rich and magnetised, providing the opportunity to study ALPs.~We now proceed to focus on bounds on $g_\mathrm{a\gamma}$ inferred from X-ray observations of AGN hosted by galaxy clusters.~Clusters are permeated by the intracluster medium (ICM), known to be magnetized up to $\sim$$10 \ \mu\mathrm{G}$ within cool-core cluster cores.~While the 3-dimensional structure of the ICM field cannot directly be mapped, the strength and typical coherence length of the ICM field can be inferred from measurements of Faraday Rotation Measures (RM) at radio wavelengths.~RMs have also indicated that ICM fields are not regular but turbulent on scales $100 \ \mathrm{pc} - 10 \ \mathrm{kpc}$ \cite[][]{govoni_2012_summary_bfields_icm}.~

In many clusters, the central brightest cluster galaxy (BCG) hosts an X-ray luminous AGN.~As the X-rays from the AGN traverse the magnetised ICM, photon-ALP interconversion would imprint distortions on the intrinsic AGN spectrum.~The absence of observed distortions can be used to exclude regions of the $(m_\mathrm{a},g_\mathrm{a\gamma})$ parameter space for $m_\mathrm{a}<10^{-12} \ \mathrm{eV}$.~The current most sensitive bounds on these very-light ALPs were inferred with this technique using \textit{Chandra} Transmission Grating Spectroscopy of the luminous cluster-hosted AGN H1821$+$643 \citep[][]{sisk21_alps} (Tab.~\ref{table:t0_alpBounds_current_andNominal_nextGen}).~We refer to \citet{Wouters_FirstConstraints, Marsh17, Conlon_17_manySources, Berg:2016_NGC1975} for previous astrophysical ALP bounds inferred from X-ray observations of AGN hosted by or located behind clusters.~The purpose of this paper is to assess the capabilities of future X-ray missions, particularly the \textit{Athena} and \textit{AXIS} observatories, to probe ALP physics.

\begin{table*}
\centering 
\renewcommand{\arraystretch}{1.2}
\begin{tabular}{c c c c}
\hline \hline
  Source & Instrument & Current bounds on $g_\mathrm{a\gamma}$ (\textit{not} excluded) at the $95\% \ \mathrm{level}$ & Reference \\ \hline \hline
  NGC1275 & \textit{Chandra}/HETG & $g_\mathrm{a\gamma} < 4.0 \times 10^{-13} \ {\mathrm{GeV}}^{-1}$ & \citet[][Model B]{Reynolds20} \\
  A1795Sy1$^{\dagger}$ & \textit{Chandra} & $g_\mathrm{a\gamma} \lesssim 0.6 \times 10^{-12} \ {\mathrm{GeV}}^{-1}$ & \citet[][]{Schallmoser2021_updated_ML} + ML \\
  H1821+643 & \textit{Chandra}/LETG+HETG & $g_\mathrm{a\gamma} < 5.0 \times 10^{-13} \ {\mathrm{GeV}}^{-1}$ & \citet[][]{sisk21_alps} \\ \hline \hline 
 \multicolumn{4}{c}{Projected bounds $g_\mathrm{a\gamma}$ (\textit{not} excluded) at the $95\%$ level from next-generation X-ray telescopes} \\ \hline \hline 
 NGC1275 & \textit{Athena}/X-IFU$^{*,\dagger}$ & $g_\mathrm{a\gamma} < 6.3 \times 10^{-14} \ {\mathrm{GeV}}^{-1}$ & This work, no calibration (+ML) \\
  NGC1275 & \textit{Athena}/X-IFU$^{\dagger}$ & $g_\mathrm{a\gamma} < 2.0 \times 10^{-13} \ {\mathrm{GeV}}^{-1}$ & This work, inc.~calibration with ML \\
  NGC1275 & \textit{AXIS}/on-axis$^{*}$ & $g_\mathrm{a\gamma} < 2.0 \times 10^{-13} \ {\mathrm{GeV}}^{-1}$ & This work, no calibration \\
\hline \hline
\end{tabular}
\caption{Top: List of current most sensitive bounds on very-light ALPs ($m_\mathrm{a} < 10^{-12} \ \mathrm{eV}$) based on single-source spectroscopy studies (see references in Sec.~\ref{section:s1_into}).~\citet[][]{Schallmoser2021_updated_ML} used a 3-dimensional field model and machine learning classifiers$^{\dagger}$ to improve on the previous bound from A1795Sy1 found by \citet[][]{Conlon_17_manySources}.~Bottom: projected bounds on very-light ALPs from the next-generation X-ray telescopes \textit{Athena}/X-IFU and \textit{AXIS} (when used for on-axis observations), based on their optimal calibration$^{*}$ (see Sec.~\ref{section:s5_nominal_nextGen}).~The upper bound on $g_\mathrm{a\gamma}$ quoted for \textit{Athena}/X-IFU under the assessment of detector calibration$^{\dagger}$ is found by a Deep Neural Network trained on simulated spectra of NGC1275 with injected ALPs at different $g_\mathrm{a\gamma}$ across a library of conservative detector responses (see Sec.~\ref{section:s7_CNN_ML}).}
\label{table:t0_alpBounds_current_andNominal_nextGen}
\end{table*}

The \textit{Athena X-ray Observatory} is the European Space Agency's next flagship X-ray observatory and will offer spatially-resolved spectroscopic capabilities exceeding those of current and next-generation missions expected to launch in the near future \citep{barret_latest}.~The leap in the science \textit{Athena} will provide is due to:~the unprecedented spectral resolution ($2.5 \ \mathrm{eV}$) of its X-ray Integral Field Unit (the \textit{Athena}/X-IFU or X-IFU); \textit{Athena}'s large effective area (see Tab.~\ref{table:t1_athena_comp_axis_chandra}); and an arc-sec-scale spatial resolution across a broadband coverage \citep[$0.5-12 \ \mathrm{keV}$; see][]{Barret2020_AthenaDescription,barret_latest}.~These factors provide \textit{Athena} with exciting prospects for cluster-ALP searches, given that an X-ray telescope with a fine energy resolution such as the X-IFU will be able to discriminate further amongst photon-ALP mixing models suited to describe the residuals in high-quality AGN spectra.

The capabilities of undertaking ALP studies with \textit{Athena} under optimal detector calibration were first studied by \citet[][]{Conlon_2017_ProjectedBoundsOnAthena}, who considered the projected bounds on very-light ALPs from a simulated observation of the central AGN in the Perseus cluster, NGC1275, with $200$-$\mathrm{ks}$ of exposure.~They found projected bounds on very-light ALPs of $g_\mathrm{a\gamma} < 1.5\times 10^{-13} \ {\mathrm{GeV}}^{-1}$ at $95\%$ confidence.

In this paper, we revisit the potential of \textit{Athena} to constrain ALPs by extending the previous considerations of \citet[][]{Conlon_2017_ProjectedBoundsOnAthena} to include the effects of instrumental miscalibration.~Experience with all previous X-ray spectroscopes
has highlighted the difficulty of obtaining (relative) calibration errors below the 1–2\% level.~Importantly, the performance requirements of the X-IFU are subject to knowing both the broadband shape and normalisation (within 0.5-10 keV and at 1 keV, respectively) of the effective area curve with above 97\% precision.~We refer to Sec.~5 of \citet[][]{Barret_and_cappi_2019_bhspin_degen} and Sec.~2 of \citet[][]{barret_latest} for detailed descriptions of these requirements, which imply that any high-quality spectrum of a sufficiently bright AGN seen by \textit{Athena}/X-IFU will be dominated by systematic uncertainties, namely detector miscalibration.~Generically, this would not apply to high-quality AGN spectra inferred by \textit{Chandra} or \textit{XMM-Newton} since these would be dominated by statistical uncertainties due to their smaller correcting area.

Using a library of realistic (conservative) \textit{Athena}/X-IFU detector responses \citep[such as in][]{Barret_and_cappi_2019_bhspin_degen}, we show that the application of the standard analysis techniques in the presence of calibration errors hobbles the ALP constraints from {\it Athena}/X-IFU data.~However, spectral features resulting from instrumental miscalibration are generally quite distinct from those induced by ALP-photon mixing.~We show that the application of machine learning techniques can, to a very large extent, circumvent calibration uncertainties when evaluating \textit{Athena}'s projected bounds on ALPs.~We conclude that the use of such techniques may be key in unveiling the potential of other next-generation X-ray observatories to further probing ALP parameter space.

We also present the first projected bounds on ALPs from the \textit{Advanced X-ray Imaging Satellite} (\textit{AXIS}), a probe-class concept that will be proposed to NASA for a 2032 launch.~While {\it AXIS} has a smaller collecting area and significantly poorer spectral resolution than {\it Athena}, its superior spatial resolution of $1\arcsec~$half-power diameter (HPD) offers its own advantages for studying cluster-embedded AGN.~

This paper is organised as follows.~Sec.~\ref{section:s2_athena_andaxis} summarises the \textit{Athena} and \textit{AXIS} missions.~We proceed by simulating a $200$-$\mathrm{ks}$ \textit{Athena}/X-IFU and \textit{AXIS} observation of NGC1275 under the assumption of optimal detector responses (i.e.~optimal calibration, Sec.~\ref{section:s3_simulated_spectra_ngc1275_nominal}).~The grid of photon-ALP mixing models employed in our analysis pipeline is described in Sec.~\ref{section:s4_alpGrid}.~The projected bounds on very-light ALPs from such missions are presented in Sec \ref{section:s5_nominal_nextGen}.~In Sec.~\ref{section:s6_bounds_differentCalibration}, we address ALP constraints with \textit{Athena} in the presence of realistic calibration errors by marginalising over a set of non-optimal detector responses.~In Sec.~\ref{section:s7_CNN_ML}, we introduce the exciting prospect of using machine learning techniques --in particular, deep neural networks-- to disentangle the effects of detector calibration and photon-ALP signatures for \textit{Athena} and next-generation X-ray observatories.~In Sec.~\ref{section:s7_injSignal}, we present the projected recovery of several ``injected'' ALP signals by \textit{Athena}/X-IFU under the effects of detector calibration and magnetic field model.~We discuss our results and limitations and conclude in~Secs.~\ref{section:s9_discussions} and~\ref{section:s10_conclusions}, respectively.

\section{The \textit{Athena} and \textit{AXIS} observatories}
\label{section:s2_athena_andaxis}

The \textit{Athena X-ray Observatory} is the second L(large)-class mission selected by the European Space Agency within the Cosmic Vision Program to address the {\it Hot and Energetic Universe} scientific theme \citep[][]{n2013hot_AthenaWhitePaper}.~In this work, we consider the baseline \textit{Athena} design and capabilities as described by \citet[][]{Barret2020_AthenaDescription}.~\textit{Athena} will consist of a single X-ray telescope constructed from iridium-coated silicon pore optics with a focal length of $12 \ \mathrm{m}$, permitting imaging with a Half Energy Width (HEW) of 5\,arcsec and a mirror effective area at 1\,keV of 1.4 m$^2$.~A hexapod mechanism allows the focused X-ray beam to be directed at one of two focal plane instruments:~the Wide Field Imager (WFI), consisting of an array of Active Pixel Sensors based of DePFET; and the X-ray Integral Field Unit (X-IFU), consisting of a cryogenically cooled microcalorimeter.~Given its unprecedented spectral resolution and effective area (Tab.~\ref{table:t1_athena_comp_axis_chandra}), the X-IFU will provide the high-quality view of the intrinsic AGN emission and nearby ICM needed for sensitive ALP constraints.

{\it AXIS} is a response to the NASA's {\it Astrophysics Probe Explorer Program} and answers the call by the Astro2020 Decadal Survey for a probe-class X-ray or mid-Infrared Satellite to launch in 2032 \citep[][]{pathwaysToDiscovery_2021}.~\textit{AXIS} will be a high-spatial resolution X-ray observatory that provides powerful synergy and complementarity with other facilities expected in the 2030s, including {\it Athena}.

{\it AXIS} achieves 1\,arcsec HPD image quality across a $24\times24 \ \mathrm{arcmin}^2$~field of view and a mirror effective area of 7\,000\,cm$^2~$via a novel mirror design that employs precision cut mono-crystaline silicon foils.~The detector consists of a cooled fast-readout Digital CCD, and is protected from both optical light and molecular contamination by a set of filters.~Including the filter transmission curves and the detectors' quantum efficiency, the effective area of the observatory is $5\,700$ cm$^2$ at 1\,keV.~That is,~an order of magnitude greater than {\it Chandra} when used in imaging mode with the {\it Advanced CCD Imaging Spectrometer} (ACIS).

\begin{figure}
 \center
 \includegraphics[width=0.4\textwidth]{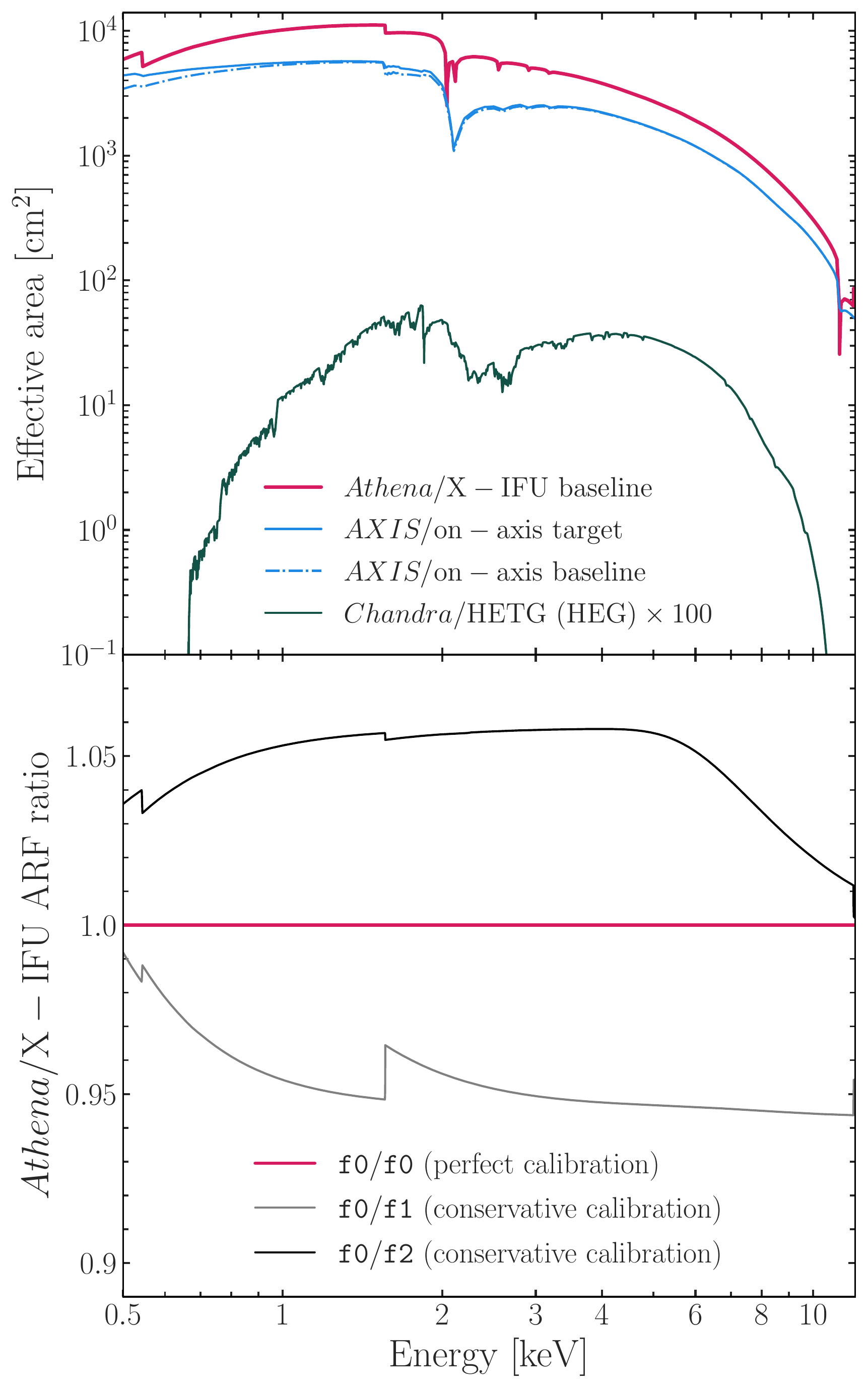}
 \centering
 \caption{Top panel: optimal (baseline) effective area responses of \textit{Athena}/X-IFU, \textit{Chandra}/HETG (when its High-Energy Grating is read on ACIS-S), and \textit{AXIS} (for on-axis observations under its baseline and target requirements).~Bottom panel: ratio of the effective areas of \textit{Athena}/X-IFU's optimal response (in magenta) divided by two realistic conservative responses (\texttt{f1}, \texttt{f2}).~We refer to \citet[][]{Barret_and_cappi_2019_bhspin_degen} for a description on how the latter were produced.}
 \label{figure:f1_effectiveArea_threeMissions}

\end{figure}

The energy-dependent effective areas of \textit{Athena}/X-IFU, \textit{Chandra}/HETG and \textit{AXIS} are shown by Fig.~\ref{figure:f1_effectiveArea_threeMissions}.~All responses present a narrow drop at $1.5-2.0 \ \mathrm{keV}$, corresponding to the absorption edge of the material coating the mirror.~For the baseline ARF scenario, \textit{AXIS}'s effective area (for on-axis observations of point-like sources) would differ from that in the target scenario due to a limited performance at energies below $1 \ \mathrm{keV}$ due to a significantly thicker optical blocking filter.

\begin{table*}
\centering 
\renewcommand{\arraystretch}{1.2}
\begin{tabular}{c c c c c}
\hline
  X-ray mission & Spectral resolution & Angular resolution & Collecting area at $1 \ \mathrm{keV}$ & Broadband energy range\\
  \hline 
  \textit{Athena} & ${2.5 \ \mathrm{eV}}^{\dagger} \ \mathrm{at } \ 1 \ \mathrm{keV}$ & $5 \ \mathrm{arcsec}$ & $1 \ \mathrm{m^{2}}$ & $0.5 - 12 \ \mathrm{keV}$ \\ 
  \textit{AXIS} & $150 \ \mathrm{eV \ at} \ 6 \ \mathrm{keV}$ & $1 \ \mathrm{arcsec}$ & $0.56 \ \mathrm{m^{2}}$ & $0.2 - 12 \ \mathrm{keV}$ \\ 
   \textit{Chandra}/HETG & $150 \ \mathrm{eV}$ at $1 \ \mathrm{keV}$ & $0.5 \ \mathrm{arcsec}$ & $0.04 \ \mathrm{m^{2}}$ & $2.0 - 12 \ \mathrm{keV}$ \\ 
\hline
\end{tabular}
\caption{Overview of the designs of \textit{Athena}, \textit{AXIS} (for its average FOV response) and \textit{Chandra}/HETG \citep[quoted for its High Energy Grating, HEG, after the first five years of performance -- see][]{canizares_chandra_hetg_after5years}.~All \textit{Athena} specifications result from the combination of the capabilities of both of its instruments, the WFI and X-IFU, unless flagged with a dagger ($^\dagger$) symbol, in which case they correspond to those of X-IFU only.}
\label{table:t1_athena_comp_axis_chandra}
\end{table*}  \section{Simulating {\it Athena} and {\it AXIS} observations of NGC1275}
\label{section:s3_simulated_spectra_ngc1275_nominal} 
NGC1275 is the central BCG of the Perseus cluster, the most massive \citep[$M_{200} = (6.6 \pm 0.4)\times 10^{14} M_\odot$,][]{Simionescu2012_gasmotions_perseus} nearby cluster \citep[at redshift $z = 0.017284$, ][]{2018_hitomi_citation_for_z} whose ICM structure and dynamics have been extensively studied over the years with numerous X-ray missions \citep[e.g.][]{DeepChandra_Perseus_Fabian06, Simionescu2012_gasmotions_perseus, Zhuravleva2014_turbulentPerseusVirgo}.~NGC1275 hosts a luminous AGN that is a prime target for ALP searches.~

We refer to \citet[][]{2021_reynolds_circumnuclearEnv_ngc1275} for a recent X-ray study of this AGN and its circumnuclear environment.~\citet[][]{2021_reynolds_circumnuclearEnv_ngc1275} employed the same $490$-$\mathrm{ks}$ \textit{Chandra}/HETG observation previously used to set tight bounds on ALPs \citep[see Tab.~\ref{table:t0_alpBounds_current_andNominal_nextGen} and][]{Reynolds20}.~Formerly, the study of Perseus and NGC1275 had been the first science target for the \textit{Hitomi} Soft X-ray Spectrometer, an X-ray microcalorimeter which provided unprecedented details on the dynamics and thermodynamics of the ICM emission as well as the Fe-$K\alpha$ region within the core of NGC1275 \citep[][]{Hitomi_Perseus_mainpaper2016,2018_hitomi_resolved_fekalpha_line,atomicdata_hitomi_2018}.~

Below, we present the spectral models we assumed to simulate \textit{Athena}/X-IFU and \textit{AXIS} observations of NGC1275.~Throughout the entirety of our analysis, we assume a flat ($\Omega_\mathrm{\kappa} = 0.0$) Universe compatible with $\Lambda$CDM cosmology, with: $H_0 = 70 \ \mathrm{km/s/Mpc}$, $\Omega_\mathrm{m} = 0.3$, $\Omega_\mathrm{\Lambda} = 0.7$ \citep[][]{PlanckCollaborationLatestResults18}.~At a redshift of $z = 0.017284$, this places NGC1275 at a luminosity distance of $75 \ \mathrm{Mpc}$.~We assume a hydrogen column density local to the Milky Way of $N_\mathrm{H} = 1.32 \times 10^{21} \ \mathrm{cm^{-2}}$ \citep[][]{kaberla} and use the element abundance ratios of \citet[][]{Wilms_2000_abundances}.

\subsection{The baseline model}
\label{subsection:s3.1_simulated_xifu_axis}
\begin{figure}
 \center
 \includegraphics[width=0.45\textwidth]{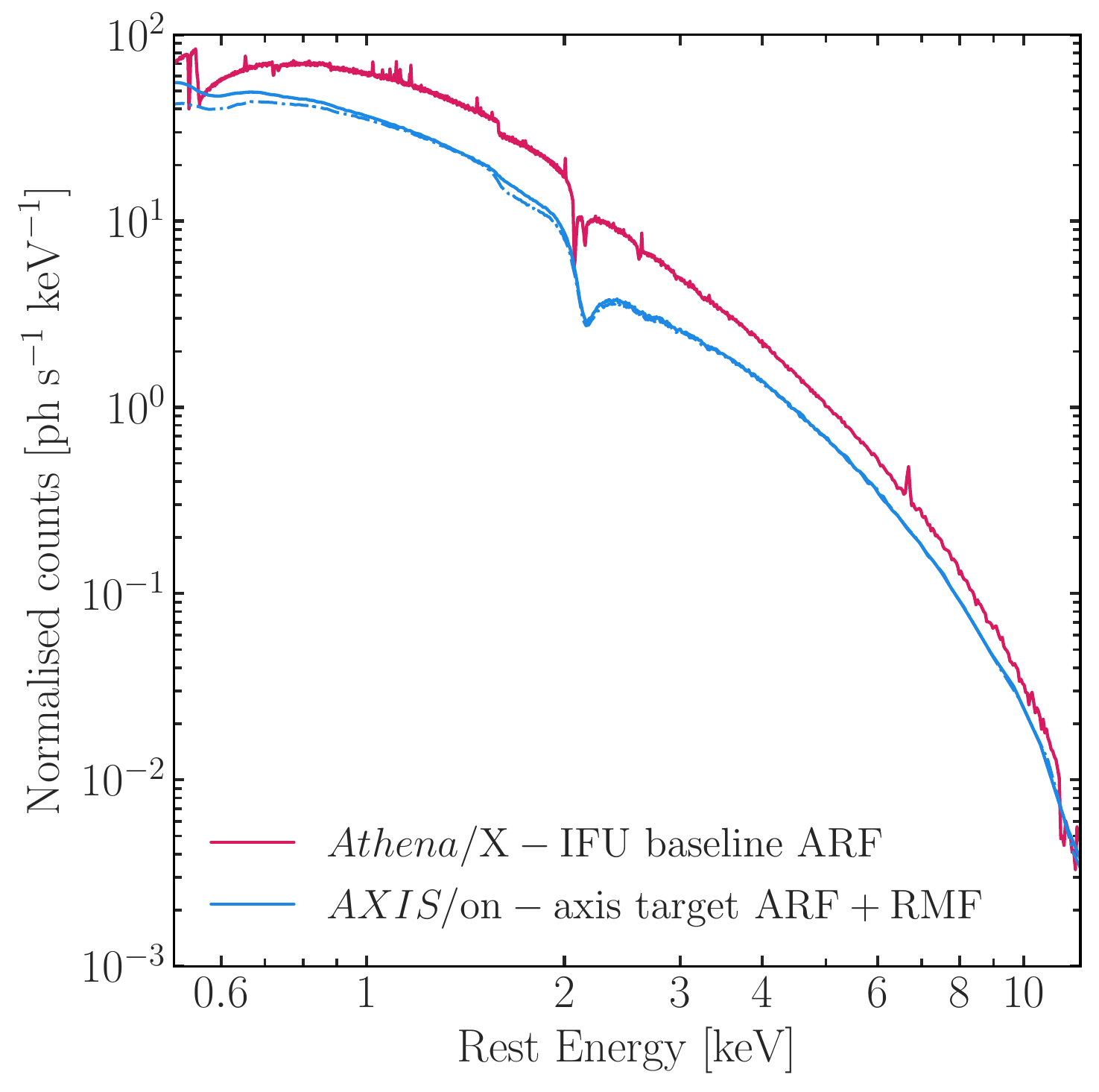}
 \caption{Effective-area corrected simulated spectra of NGC1275 seen by \textit{Athena}/X-IFU and \textit{AXIS} (used for on-axis observations under its target ARF+RMF specfications) when read on their optimal detector responses (ARFs).~\textit{AXIS}'s baseline and target RMFs differ only due to the improved spectral energy resolution of the latter at lower energies.~The dash-dotted line shows the predicted count rate for a $200$-$\mathrm{ks}$ \textit{AXIS}/on-axis observation of NGC1275 under its baseline ARF+RMF requirements.~For plotting purposes, an energy binning scheme of a target signal-to-noise ratio of 100 and 25, with the restriction that no more than 200 and 75 spectral bins were coadded, was applied to the \textit{Athena}/X-IFU and \textit{AXIS} spectra, respectively.}
 \label{figure:f2_simulated_xifu_axis_chandra}
\end{figure}

Using the~$490$-$\mathrm{ks}$~\textit{Chandra}/HETG observation of NGC1275, \citet[][]{Reynolds20} found the AGN to be well characterised by an unabsorbed power-law with photon index $\Gamma\approx 1.9$ modified by the effects of Galactic absorption, yielding residuals only below the $3\%$ level outside of the iron band.~In \citet[][]{2021_reynolds_circumnuclearEnv_ngc1275}, the description of the spectrum was found to improve slightly with the use of a partial-covering intrinsic absorber, interpreted as a composite X-ray source with an absorbed jet working surface and an unabsorbed accretion disk corona.~This interpretation also brings the X-ray data in line with the detection of molecular absorption in \textit{ALMA} observations \citep[][]{Nagai_2019_ngc1275}.

However, \citet[][]{2022_matthews_bfieldModels} found that the upper bound on $g_\mathrm{a\gamma}$ inferred from \textit{Chandra} data of NGC1275 (see Tab.~\ref{table:t0_alpBounds_current_andNominal_nextGen}) remains insensitive to whether a partial covering absorber is accounted for in the spectral fits to the AGN spectrum.~Thus, for simplicity, in this work we adopt the simple power-law model in order to generate simulated (fake) \textit{Athena} and \textit{AXIS} spectra of NGC1275.~These simulated spectra were produced with \textsc{PyXspec}'s \textit{fakeit} command under the consideration of Poisson noise statistics and neglecting background emission.~In all cases, we consider an on-source exposure of $200$-$\mathrm{ks}$ for comparison with \citet[][]{Conlon_2017_ProjectedBoundsOnAthena}.~

\subsection{Simulated \textit{AXIS} observations of NGC1275}
\label{subsubsection:axis_sim_spectra_nominal}
Given \textit{AXIS}'s superior spatial resolution (see Tab.~\ref{table:t1_athena_comp_axis_chandra}), we assume that it can capture a spectrum of NGC1275 free from the contributions of the surrounding ICM.~Thus, the baseline model for our {\it AXIS} simulations is just an absorbed power-law (\texttt{tbabs * po}); see Fig.~\ref{figure:f2_simulated_xifu_axis_chandra}.~We note that we have not included the influence of photon pileup in our simulated \textit{AXIS} spectra, as follows.~In this work, we operate under the assumption that {\it AXIS} will have bright-source modes which permit accurate CCD spectroscopy of sources with fluxes at least as high as a few\,milliCrab.

\subsection{Simulated \textit{Athena}/X-IFU observation of NGC1275}
\label{subsubsection:athena_xifu_sim_spectra_nominal}
Given an angular resolution of~5$\arcsec$~HEW, an \textit{Athena}/X-IFU observation of a cluster-hosted AGN will necessarily suffer from ICM contamination.~However, excellent spectral separation of the AGN and ICM emission should be possible given the combination of the \textit{Athena}/X-IFU effective area and spectral resolution.~

We simulate an \textit{Athena}/X-IFU observation of NCG1275 using the \texttt{bvvapec} single-temperature plasma model to model the velocity broadened ICM emission lines.~All model parameters were taken from the reference \texttt{apec-v.3.0.8} model that had been fitted to \citet[data from][see their Tab.~1]{atomicdata_hitomi_2018}, with the exception of its normalisation parameter ($A_\mathrm{plasma}$ in Tab.~\ref{table:t2_astroModel_specifications_baseline}), for reasons specified below.~The element abundances within \texttt{bvvapec} of the elements \textit{not} listed in Tab.~1 of \citet[][]{atomicdata_hitomi_2018} were assumed to be $1.0$ relative to the proto-solar metallicities of \citet[][]{loders_palmer_protosolar_abund_apec}.

Given \textit{Hitomi}'s~0.5$\arcmin$~angular resolution, the normalisation of the \texttt{apec-v3.0.8} model fitted to \citet[][]{atomicdata_hitomi_2018} modelled the ICM emission across a large volume within Perseus.~Considering \textit{Athena}'s improved angular resolution compared to \textit{Hitomi}'s, a suitable value of $A_\mathrm{plasma}$ for our analysis was found as follows.

We firstly identified the single longest observation of NGC1275 taken with \textit{Chandra}.~This observation (ObsID: 4952; exposure $164.24$-$\mathrm{ks}$) had started in October 2004 and had been read on the ACIS-S detector.~We reprocessed these data with the \textsc{CIAO} software\footnote{See \hyperlink{}{https://cxc.cfa.harvard.edu/ciao/}} v.4.14.~We extracted a spectrum of the ICM emission from an annuli centered on NGC1275 with inner/outer radii of 1$\arcsec$/10$\arcsec$ and 2$\arcsec$/10$\arcsec$ in an attempt to isolate the ICM emission in a plausible {\it Athena} extraction region.~To first order, these spectra were found to be well-described by an \texttt{apec} model with a normalisation component of $2\times 10^{-3} \ {\mathrm{cm}}^{-3}$.~We therefore adopted the value of $A_\mathrm{plasma} = 2\times 10^{-3} \ {\mathrm{cm}}^{-3}$ in our simulated \textit{Athena}/X-IFU spectrum of NGC1275 (see Tab.~\ref{table:t2_astroModel_specifications_baseline}).

Our simulated \textit{Athena}/X-IFU spectra of NGC1275 were generated using the spectral model consisting of this ICM component added to the absorbed power-law model, that is, via \texttt{tbabs * (po + bvvapec)}.~The resulting \textit{Athena}/X-IFU view of NGC1275 is shown by Fig.~\ref{figure:f2_simulated_xifu_axis_chandra}, where the narrow emission lines arising from the cluster ICM are immediately apparent when compared to the \textit{AXIS} view of the AGN.~

\begin{table*}[t]
\centering 
\renewcommand{\arraystretch}{1.2}
\begin{tabular}{c c c c c}
\hline \hline 
  \multicolumn{5}{c}{Model parameters: simulated spectra of NGC1275 under optimal ARFs}\\ \hline \hline 
  Component & Parameter & Description & \textit{Athena}/X-IFU & \textit{AXIS} \\
  \hline \hline 
  \texttt{tbabs} & $N_\mathrm{H}$ & ISM column density & \multicolumn{2}{c}{$1.32 \times 10^{21} \ {\mathrm{cm}}^{-2}$}\\ 
   \texttt{pow} & $\Gamma_X$ & Photon Index & \multicolumn{2}{c}{$1.89$} \\
  \texttt{pow} & $A_X$ & Power-law normalisation at $1 \ \mathrm{keV}$ & \multicolumn{2}{c}{$8.28 \times 10^{-3} \ \mathrm{phot ~ s^{-1} ~ {keV}^{-1}}$} \\
  \texttt{bvvapec} & $kT_\mathrm{plasma}$ & Plasma temperature & $3.969 \ \mathrm{keV}$ & - \\
  \texttt{bvvapec} & $\sigma_\mathrm{broad}$ & Doppler velocity broadening & $156 \ \mathrm{km/s}$ & - \\
  \texttt{bvvapec} & $A_\mathrm{plasma}$ & Plasma normalisation & $2\times 10^{-3} \ \mathrm{{cm}^{-3}}$ & - \\
   \texttt{bvvapec} & $z$ & Redshift & $0.017284$ & - \\
  \texttt{bvvapec} & $Z_\mathrm{Si}, Z_\mathrm{S}, Z_\mathrm{Ar}, Z_\mathrm{Ca}, Z_\mathrm{Mn}, Z_\mathrm{Fe}, Z_\mathrm{Ni}$ & Element abundances & $\neq Z_\odot$ & - \\
   \texttt{bvvapec} & $Z_\mathrm{other}$ & Element abundances & $Z_\odot$ & - \\
\hline \hline 
 \multicolumn{5}{c}{Free parameters in fits to simulated spectra of NGC1275}\\ \hline \hline 
   \texttt{pow} & $\Gamma_X$ & Photon Index & $\checkmark$ & $\checkmark$ \\
  \texttt{pow} & $A_X$ & Power-law normalisation at $1 \ \mathrm{keV}$ & $\checkmark$ & $\checkmark$ \\
  \texttt{bvvapec} & $A_\mathrm{plasma}$ & Plasma normalisation & $\checkmark$ & - \\ \hline 
\end{tabular}
\caption{Top: Model parameters used to fit the optimal simulated \textit{AXIS} (Sec.~\ref{subsubsection:axis_sim_spectra_nominal}) and \textit{Athena}/X-IFU (Sec.~\ref{subsubsection:athena_xifu_sim_spectra_nominal}) observations of NGC1275.~The \texttt{tbabs}, \texttt{po} and \texttt{bvvapec} parameter values were adopted from: \citet[][]{kaberla}; \citet[][]{Reynolds20}; and \citet[][see their Tab.~1]{atomicdata_hitomi_2018}, respectively.~Bottom: layout of the free parameters of the ALP-containing and fiducial astrophysical models simulated spectra of NGC1275.}
\label{table:t2_astroModel_specifications_baseline}
\end{table*}

\section{Grid of photon-ALP mixing models}
\label{section:s4_alpGrid}
In order to find projected bounds on ALPs from \textit{Athena} and \textit{AXIS}, we generate a library of photon-ALP mixing models for photons propagating from NGC1275 out to the virial radius of the Perseus cluster, estimated to be $R_{200} = 1.8 \ \mathrm{Mpc}$ \citep[][]{Simionescu2012_gasmotions_perseus}.~

To produce the photon-ALP mixing models, we must assume a magnetic field structure.~We use a cell-based approach to model the turbulent cluster field, specifically adopting `Model B' from \citet[][]{Reynolds20} to determine the radial profile of the cluster magnetic field and distribution of cell-sizes (``coherence lengths'').~This model is chosen so as to reproduce a value of thermal-to-magnetic pressure ratio of $\beta_\mathrm{plasma} = 100$ up to $1.8 \ \mathrm{Mpc}$ from the cluster centre.~`Model B' from \citet[][]{Reynolds20} broadly reproduces the Faraday Rotation Measures within Perseus presented by \citet[][]{Taylor06}, where the cluster acts as the Faraday screen.~Having been motivated by Rotation Measure observations of cool-core clusters \citep[][]{Taylor06, Bohringer2016_obs_betaValue}, the value $\beta_\mathrm{plasma}=100$ has been adopted in previous cluster-ALP searches \citep[][]{Reynolds20, sisk21_alps, 2022_matthews_bfieldModels}.~The derived constraints on $g_\mathrm{a\gamma}$ can readily be scaled to other choices of $\beta_\mathrm{plasma}$ \citep[see Sec 5.1 of][]{sisk21_alps}.~We also refer the reader to \citet[][]{2022_matthews_bfieldModels} for a thorough exploration of the effects of different assumptions turbulent magnetic field structure (e.g.~Gaussian Random Fields (GRFs) and ``cell-based'' models) on cluster-ALP bounds.

We generate 216 realisations of the turbulent ICM field, where each realisation reproduces a cluster field profile, $B(r)$, consistent with the notion that $\beta_\mathrm{plasma} = 100$ up to the virial radius.~Under this assumption, the \citet[][]{DeepChandra_Perseus_Fabian06} pressure profile within Perseus yields an expected magnetic field strength of $6.5 \mu \mathrm{G}$ at $25 \ \mathrm{kpc}$ from the cluster centre.~The expected field profile up to $1.8 \ \mathrm{Mpc}$ from the inner core is found by approximating the cluster as isothermal.~The field strength at comoving distance $r$ to the cluster centre \citep[shown in Fig.~4~of][]{sisk21_alps} is therefore given by $B(r) \sim 1/\sqrt{n_\mathrm{e}(r)}$ and normalised according to its value at $25 \ \mathrm{kpc}$ from the cluster centre.~Here, $n_\mathrm{e}(r)$ is the analytic \citet[][]{Churazov+03} electron number density profile found from fitting \textit{XMM-Newton} observations of Perseus.~

For each realisation, the cluster line-of-sight is split into coherence lengths in the $3.5 - 10 \ \mathrm{kpc}$ range under a power-law probability distribution \citep[][]{Vacca12_Abell2199_PowerSpec, Berg:2016_NGC1975}.~We highlight that, as specified in Sec.~3 of \citet[][]{Reynolds20}, their `Model B' accounted for the expected increase in field coherence lengths away from the cluster centre.~In each coherence length (or cell), the components of the magnetic field perpendicular to the line-of-sight (i.e.~the field components which mix with ALPs as given by Eq.~\ref{equation:eq1_interactionLangrangian_alps}) are set independently and randomly.~

We employ the Axion-Like PROpagation (\texttt{ALPRO}\footnote{See \hyperlink{}{{https://github.com/jhmatthews/alpro/tree/v1.1}{}}}) public-release code to construct a library of photon-ALP mixing curves for ALPs with $\mathrm{log}(m_\mathrm{a}/\mathrm{eV}) = -14.0$ across a grid of $\mathrm{log}(g_\mathrm{a\gamma}/{\mathrm{GeV}}^{-1}) \in [ -14.0, -11.0]$, in steps of $0.1 \ \mathrm{dex}$, for 216 configurations of the turbulent cluster field.~Although these curves are computed for a specific ALP mass, these models will accurately capture photon-ALP mixing for ``effectively'' massless ALPs with $\mathrm{log}(m_\mathrm{a}/\mathrm{eV}) < -12.0$.~Although the extrapolation to higher (but still very-light) ALP masses will generally depend on the underlying density structure of the host ICM, this does \textit{not} impact our results.

For each magnetic field realisation, the photon-ALP propagation problem described in Appendix \ref{section:a1_theory_photonALPmixing} 
is solved for an initially unpolarized beam propagating from the inner core to $1.8 \ \mathrm{Mpc}$ from the cluster centre, at all coherence cells within the line-of-sight.~The contribution from all cells results in a (total) energy-dependent photon ``survival'' probability $P_\mathrm{\gamma \rightarrow \gamma}$ for photons mixing with massless ALPs of a given $g_\mathrm{a\gamma}$.~All survival curves are sampled at $1.25 \ \mathrm{eV}$ resolution (half of the spectral resolution of \textit{Athena}/X-IFU) and computed across $0.1 - 15 \ \mathrm{keV}$.~The details of this calculation are outlined in Secs.~2 and 3 of \citet[][]{2022_matthews_bfieldModels}.

\begin{figure}
 \center
 \includegraphics[width=0.45\textwidth]{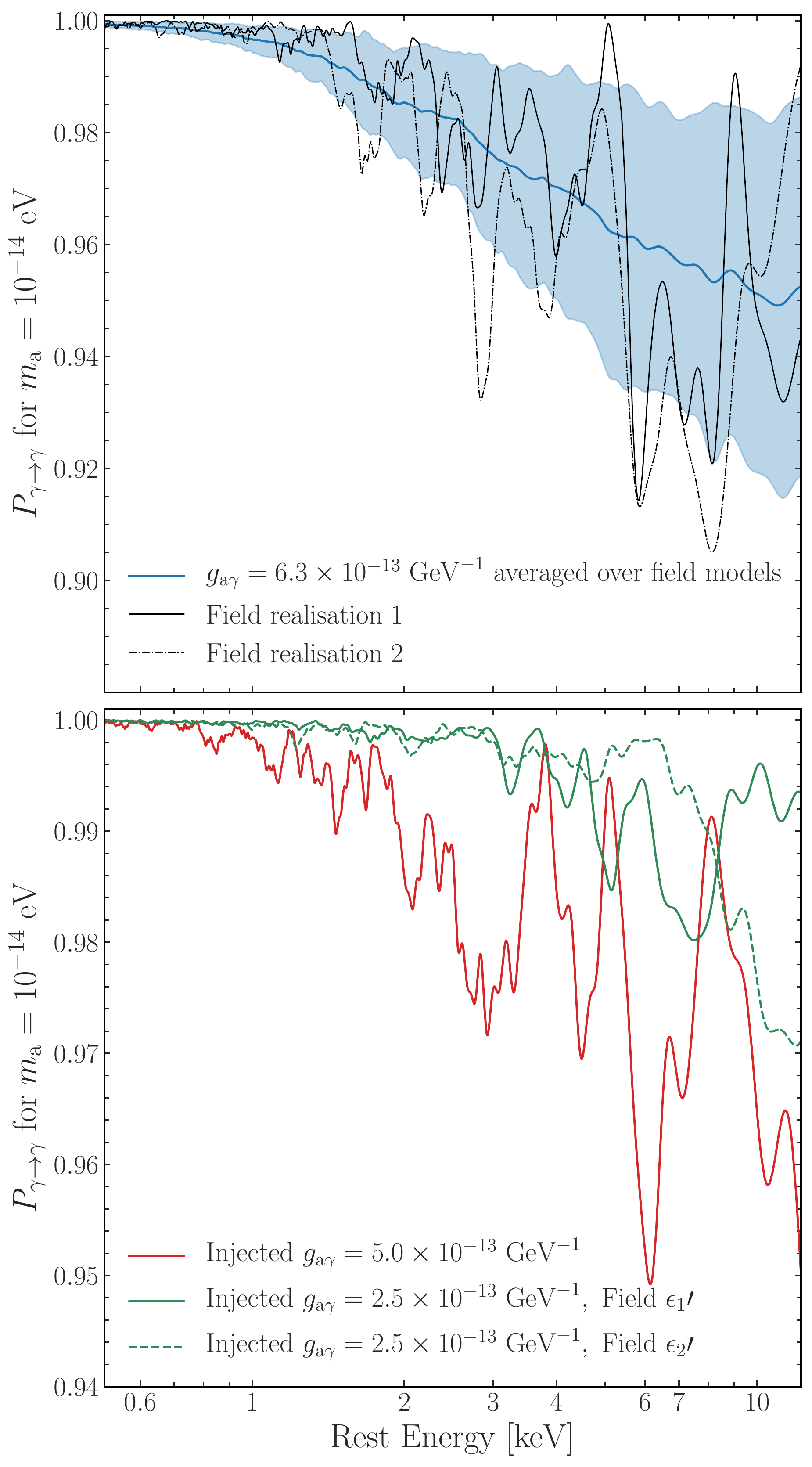}
 \caption{Top: Average photon survival probability (over 216 field realisations) for quanta propagating through the Perseus cluster when mixing with massless ALPs ($m_\mathrm{a} = 10^{-14} \ \mathrm{eV}$) of $g_\mathrm{a\gamma} = 6.3 \times 10^{-13} \ {\mathrm{GeV}}^{-1}$.~The black solid and dot-dashed lines show survival curves for two different turbulent field realisations~(referred to as $\epsilon_1, \epsilon_2$ in the text).~The shaded region shows the standard deviation of 216 curves in each energy bin, computed by using 11\,920 bins between $0.1 - 15 \ \mathrm{keV}$.~Bottom: Photon-ALP mixing models showing the injected signals in the \textit{Athena}/X-IFU analyses discussed in Sec.~\ref{section:s7_injSignal}.~The solid and dashed green curves illustrate ALP models generated under two different turbulent field models ($\epsilon_1\prime$, $\epsilon_2\prime$).}
 \label{figure:f3_mixingCurve_example_limit-currentExclusion}
\end{figure}

The survival photon-ALP mixing curves for $g_\mathrm{a\gamma} = 6.3 \times 10^{-13} \ {\mathrm{GeV}}^{-1}$ for two different configurations of the turbulent cluster field are shown by Fig.~\ref{figure:f3_mixingCurve_example_limit-currentExclusion}.~This corresponds to the upper bound on $g_\mathrm{a\gamma}$ inferred at the $99.7\%$ level by the most sensitive cluster-ALP single-source spectroscopy study to date \citep[][]{sisk21_alps}.~

\section{Optimal bounds from next-generation X-ray telescopes}
\label{section:s5_nominal_nextGen}

Equipped with the grid of photon-ALP mixing models introduced above, which we refer to as \{\texttt{ALP(i,$g_\mathrm{a\gamma}$)}\}, where~$\texttt{i}$~and~$g_\mathrm{a\gamma}$~label the field realisation and the photon-ALP coupling, respectively, we first aim to find projected bounds on ALPs from \textit{Athena} and \textit{AXIS} under the assumption that the energy-dependent area of the detectors were perfectly known.~We highlight that \textit{Athena}'s projected bounds on ALPs were first reported in Secs.~4 and 5 of \citet[][]{Conlon_2017_ProjectedBoundsOnAthena}, albeit upon a slightly different methodology to the one we follow (see their Sec.~4).

To derive the projected constraints on ALPs, we analyse the simulated ALP-free {\it Athena} and {\it AXIS} spectra following the procedure described in Sec.~4~of \citet[][]{sisk21_alps}.~We fit the simulated spectra in the observed energy range $0.5 - 12.0 \ \mathrm{keV}$.~For each \texttt{ALP(i,$g_\mathrm{a\gamma}$)} curve in our photon-ALP mixing grid, we fit a spectral model that includes the ALP-mixing (via \texttt{tbabs*(po + bvvapec)*ALP(i,$g_\mathrm{a\gamma}$)}) to the simulated \textit{Athena} spectrum (Sec.~\ref{subsubsection:athena_xifu_sim_spectra_nominal}), and \texttt{tbabs*po*ALP(i,$g_\mathrm{a\gamma}$)} to the (separate target and baseline ARF+RMF) \textit{AXIS} spectra.~The best-fit regime of each spectral fit was found by minimising the \textit{Cash} (\textit{C}) statistic \citep[][]{Cash1979,CashStatisticsUse_Kaastra_Astronomers}, appropriate for Poisson-distributed data.~\begin{figure*}
 \center
 \includegraphics[width=0.85\textwidth]{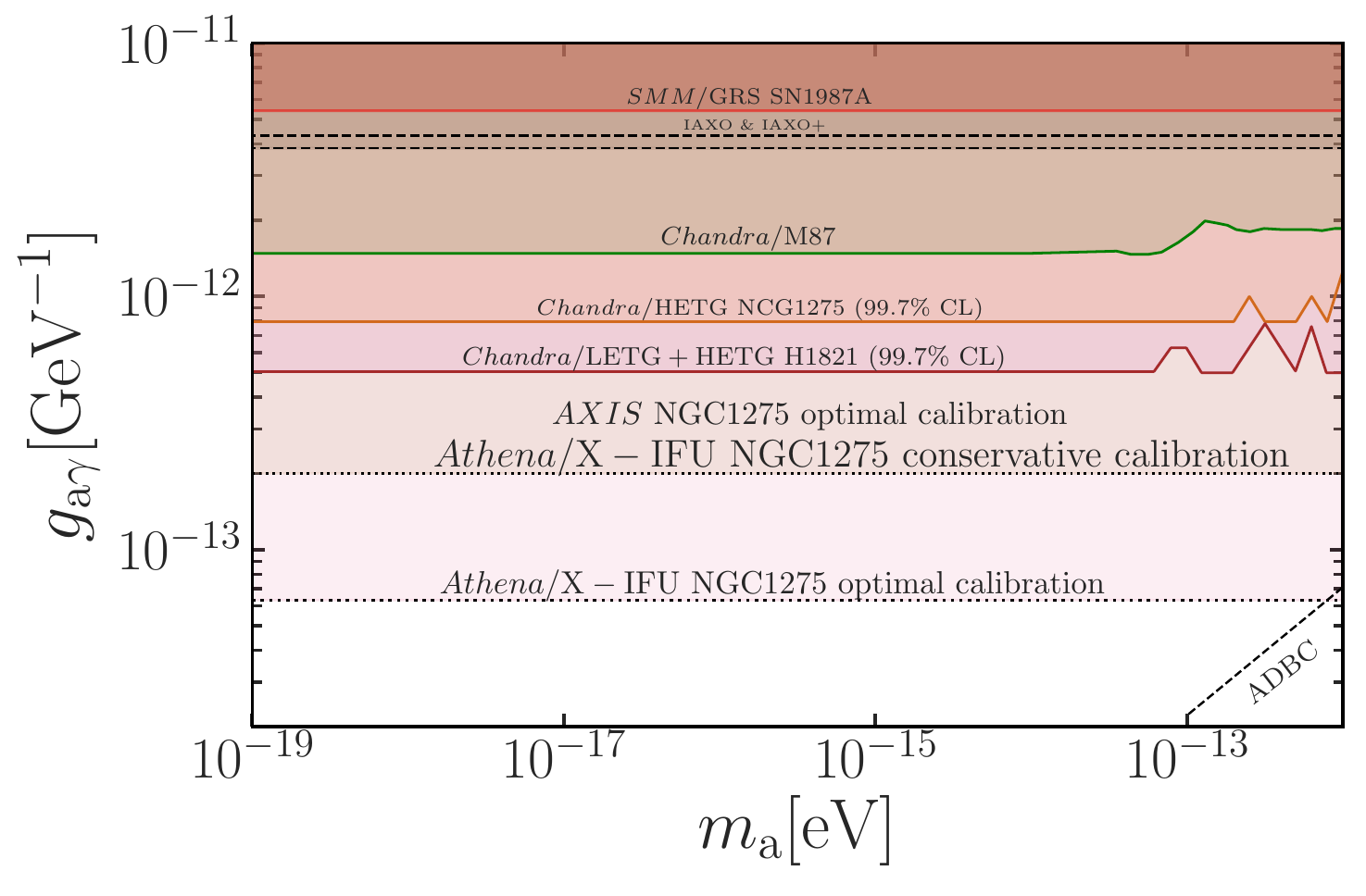}
 \caption{Projected (dotted and dashed lines) and current observational (solid lines) upper bounds (all shown at the $95\%$ exclusion level unless otherwise explicitly stated) on very-light ALPs inferred from this work and previous astrophysical searches.~We include our projected upper bounds on $g_\mathrm{a\gamma}$ from both the \textit{Athena}/X-IFU and the \textit{AXIS} observatories under the assumption of a ``perfect'' detector calibration scenario, where, in both cases, we have marginalised over 216 realisations of the turbulent field within Perseus.~We also show the projected bounds on ALPs from \textit{Athena}/X-IFU under the assessment of detector calibration (refer to Sec.~\ref{section:s7_CNN_ML}),~which overlap with the \textit{AXIS} exclusion under its optimal calibration scenario. We show the upper bounds inferred from: the SN1987A event \citep[][]{SN1987Bounds}, complemented by \textit{Fermi}/LAT observations of extra-galactic SNe \citep[][]{Meyer_2020_FermiLAT_FielfOfView_ALPs_sneExtraGalactic}; projected bounds from the next-generation axion helioscope, the International AXion Observatory (IAXO) and its upgrade \citep[IAXO+,][]{IAXO19_ProjectedSensitivity}; and existing AGN/cluster bounds from \textit{Chandra} observations of: M87/Virgo \citep[][]{Marsh17}, NGC1275/Perseus \citep[][their Model B]{Reynolds20}, and H1821+643 \citep[][]{sisk21_alps}.~The shaded regions above all curves underlie the parameter space excluded by any of the studies shown.~We include projected bounds from the future birefringent cavity experiment \textit{ADBC} (dashed black line), a future axion interferometer which will constrain ALP DM \citep[][]{adbc}.}
 \label{figure:exclusion_nominal_xifu_axis_comparison_otherAGNCluster}
\end{figure*} We find posteriors on $g_\mathrm{a\gamma}$ for $m_\mathrm{a} < 10^{-12} \ \mathrm{eV}$ by using the Bayesian framework described in Sec.~4 of \citet[][]{sisk21_alps}.~This procedure derives a normalised posterior distribution on $(m_\mathrm{a}, g_\mathrm{a\gamma}$) by marginalising over all magnetic field models.~This normalisation is achieved via Eq.~\ref{eq:eq6_normalisation_condition} and computed across $\mathrm{log}(m_\mathrm{a}/\mathrm{eV}) \in [-30.0, -12.0) \ \bigcup \ \mathrm{log}(g_\mathrm{a\gamma}/{\mathrm{GeV}}^{-1}) \in [-19.0, -10.1]$.~Operationally, all posteriors for $\mathrm{log}(m_\mathrm{a}/\mathrm{eV}) = -14.0$ are identical to those for ALPs of $\mathrm{log}(m_\mathrm{a}/\mathrm{eV}) \in [-30.0,-12.0)$.~Moreover, all $p_\mathrm{post}(m_\mathrm{a}, g_\mathrm{a\gamma})$ for $\mathrm{log}(g_\mathrm{a\gamma}/{\mathrm{GeV}}^{-1}) \in [-19.0,-14.0)$ were assigned by taking those for $g_\mathrm{a\gamma} = 1.0 \times 10^{-14} \ {\mathrm{GeV}}^{-1}$ as a proxy.~That is, in this regime, the energy-dependent modulations induced by $g_\mathrm{a\gamma}$ are indistinguishable from those induced by Poisson noise.~

Our priors on the ALP parameter space were chosen as follows.~Firstly, the mass range considered embeds the phenomenology of ``effectively'' massless ALPs, that is, where the relevant photon-ALP mixing equations (see Appendix \ref{section:a1_theory_photonALPmixing}) are mass-independent.~In general, the latter would hold for ALPs of $m_\mathrm{a} \lesssim 10^{-12} \ \mathrm{eV}$.~If such ALPs were to play a cosmological role, they \textit{could} indeed comprise dark matter.~At masses $\lesssim 10^{-30} \ \mathrm{eV}$, however, their de-Broglie wavelength will exceed the particle horizon of the Universe and ALPs will then act as dark energy \citep[][]{ALPsQuintessentialDE_Carroll}.~Secondly, for photon-ALP couplings ${g_\mathrm{a\gamma}}^{-1} \gtrsim 10^{19} \ \mathrm{GeV}$, the energy scale at which ALPs couple to electromagnetism will exceed the Planck scale, which is not expected to be realisable in consistent theories of quantum gravity.

The $95\%$~($2\sigma$)~confidence level (CL) constraints on $g_\mathrm{a\gamma}$ we infer from the simulated $200$-$\mathrm{ks}$ \textit{Athena}/X-IFU and \textit{AXIS} observations of NGC1275 under the ``perfect calibration'' scenario are listed in Tab.~\ref{table:t0_alpBounds_current_andNominal_nextGen} and shown in Fig.~\ref{figure:exclusion_nominal_xifu_axis_comparison_otherAGNCluster}.~These projected \textit{Athena} bounds significantly improve on the current best constraints (Tab.~\ref{table:t0_alpBounds_current_andNominal_nextGen}), with {\it Athena} pushing a full order of magntitude deeper than {\it Chandra}.~We refer to Sec.~\ref{section:sub9.1_summary_results_calibration_athena} for a comparison with the results in \citet[][]{Conlon_2017_ProjectedBoundsOnAthena}.~The marginalised posterior distribution (over all magnetic field realisations) for ALPs of $m_\mathrm{a} = 10^{-14} \ \mathrm{eV}$ is expected to be dominated by low values of the photon-ALP coupling,~i.e.~$g_\mathrm{a\gamma}\lesssim 6.3\times 10^{-14} \ {\mathrm{GeV}}^{-1}$ (refer to Appendix \ref{section:appendix_illustration_calibration} for discussion).

In the perfect calibration scenario, \textit{AXIS} improves upon current constraints by $0.3-0.4 \ \mathrm{dex}$ for both its baseline and target area responses.~The latter is shown in Fig.~\ref{figure:exclusion_nominal_xifu_axis_comparison_otherAGNCluster}.

\section{The effects of calibration on \textit{Athena}'s projected bounds on ALPs}
\label{section:s6_bounds_differentCalibration}
Following on from \citet[][]{Conlon_2017_ProjectedBoundsOnAthena}, we have shown that \textit{Athena}/X-IFU, with its unprecendented spectral energy resolution of $2.5\ \mathrm{eV}$ and effective area of about $1 \ \mathrm{m}^2$ at 1 keV, could push down the current most sensitive bounds on $g_\mathrm{a\gamma}$ by an order of magnitude provided we have perfect knowledge of its energy-dependent area.~However, no X-ray observatory has perfect calibration, with the practicalities of both ground and on-orbit calibration leading to irreducible uncertainties in the (relative) effective area curves at the level of 1--2\%.~In view of this reality, we now proceed to reassess \textit{Athena}'s power to constraining very-light ALPs under imperfect calibration.~The importance of such an exercise is discussed in Secs.~\ref{section:s6.1_importance_arf_illustrated} and in Appendix \ref{section:appendix_illustration_calibration}.~In Sec.~\ref{subsection:s6.2_marginalising_over_differentARFs}, we then quantify the effects of detector calibration in \textit{Athena}'s projected bounds on ALPs using a $Cash$ statistic likelihood method that extends from that employed in the ``optimal'' calibration scenario.~In Sec.~\ref{section:s7_CNN_ML}, we introduce the exciting prospect of disentangling calibration residuals and ALP features via the use of machine learning, thus leading to tighter bounds on $g_\mathrm{a\gamma}$.
 \begin{figure*}
 \center
 \includegraphics[width=1.\textwidth]{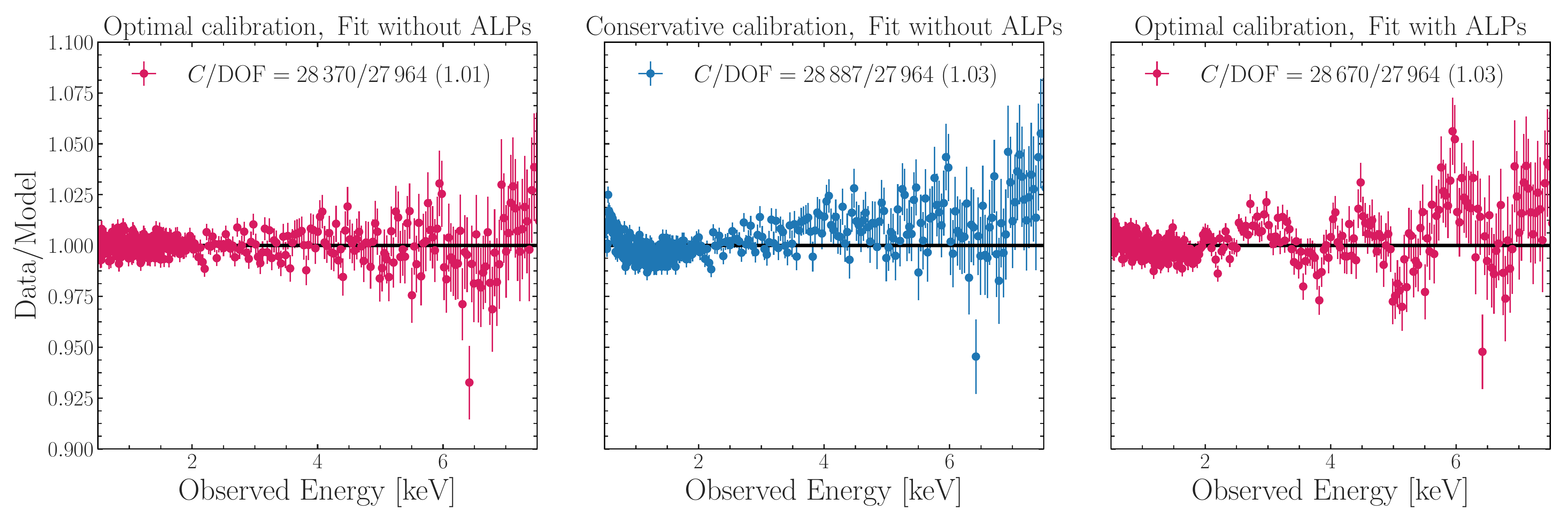}
 \caption{Residuals from fitting the synthetic \textit{Athena}/X-IFU spectrum presented in Sec.~\ref{section:s3_simulated_spectra_ngc1275_nominal} with a variety of models.~The different panels illustrate the effects of detector calibration and of photon-ALP mixing when fitting the synthetic data as follows.~Left panel:~a ``fiducial'' (astrophysics-only) model is fitted under an optimal detector calibration scenario.~Central panel: a ``fiducial'' (astrophysics-only) model is fitted to the data under a conservative calibration scenario.~Right panel: a multiplicative ALP model (featuring $g_\mathrm{a\gamma} = 5.0\times 10^{-13} \ {\mathrm{GeV}}^{-1}$) is fitted under an optimal calibration scenario.~The simulated data was fitted within the observed energies $0.5 - 12 \ \mathrm{keV}$.~For plotting purposes, a binning scheme of a target signal-to-noise ratio 250 with the restriction that no more than 100 spectral bins were coadded was applied, and we show the energy range $0.5 - 7.5 \ \mathrm{keV}$.~The best-fit $C$-stat of each fit, the number of degrees of freedom (DOF), and the reduced $C$-stat ($\equiv C/\mathrm{DOF}$) are quoted (with the latter in parenthesis).~The ALP fit corresponds to $\mathrm{log}(g_\mathrm{a\gamma}/{\mathrm{GeV}}^{-1})=-12.3$,~i.e.~close to the current exclusion on $g_\mathrm{a\gamma}$~(refer to Tab.~\ref{table:t0_alpBounds_current_andNominal_nextGen}).}
 \label{figure:f4_residuals_differentARFs}
\end{figure*}
\subsection{Overview of the effects of non-optimal calibration}
\label{section:s6.1_importance_arf_illustrated}

One may worry that the energy-dependent residuals induced by the calibration of a given detector (encapsulated in the Area Response File, ARF) could mimic the energy-dependent oscillations induced by photon-ALP mixing.~We examine this issue using a library of sampled effective area curves (ARFs) for {\it Athena}/X-IFU which was produced as described in \citet[][]{Barret_and_cappi_2019_bhspin_degen}.~These ARFs capture the realistic effects of instrumental modelling uncertainties on detector calibration.~The effects of modelling uncertainties of the mirror response on detector calibration were \textit{not} taken into account and are therefore beyond the scope of this paper.~These mirror-induced calibration effects were recently explored by \citet[][]{athena_guainazzi_calibration}.~For reference,~Fig.~\ref{figure:f1_effectiveArea_threeMissions}~illustrates a subsample of realistic ``non-optimal'' ARFs employed in our analysis.
 
Together with the energy response matrix, the ARFs relate the observed photon fluxes to the actual source spectrum.~A given detector's ARF is sensitive to multiple design features, for instance:~the level of impurity and thickness of the material coating the mirror; the orientation of the mirror,~etc.~Thereby, in general, a representative set of ARFs for next-generation X-ray observatories must be constructed in order to assess the extent to which these missions will be able to meet their science objectives.

We assess the effects of detector calibration on \textit{Athena}'s projected bounds on ALPs using the library of ARFs outlined in Sec.~5 of \citet[][]{Barret_and_cappi_2019_bhspin_degen}.~Each ARF in the library was generated so that its normalisation remained unchanged at 1 keV and throughout 0.5-10 keV compared to the nominal ARF; and subject to a change in the effective area shape of up to 3\% with respect to the nominal ARF.~As the end product, each ARF in the library captured the uncertainty in the thickness of instrument filters and absorbers without accounting for uncertainties around the edge of the response.~The latter will be considered in a future paper.

We can obtain a first estimate of the effect of calibration expected from pure Poisson statistics by assuming that the ``true'' ARF is an unknown member of the library of ARFs.~Following \citet[][]{drake_2006_chandraResponse}, one can then proceed by:~(i) either simulating a single spectrum with the nominal ARF and fitting it across all ARFs in the library; (ii) or by generating a single spectrum per ARF in the library and subsequently fitting it with the nominal ARF.~Following Sec.~5 of \citet{Barret_and_cappi_2019_bhspin_degen}, we use the former method as it is able to more closely capture the perturbative nature of detector miscalibration in the recovery of best-fit parameter values; and we refer to \citet[][]{cucchetti_18_recentEx_cal} for an implementation of the latter method.~In both approaches, the realistic effects of detector miscalibration can be inferred from a Monte Carlo approach by recording the best-fit parameter values when a spectral model is fitted to the simulated spectra.~We refer to Fig.~\ref{figure:f4_residuals_differentARFs}~and Appendix \ref{section:appendix_illustration_calibration} for a qualitative assessment on the realistic impact of calibration in \textit{Athena}'s sensitivity to constraining ALPs.


\subsection{Revisiting \textit{Athena}'s projected bounds on ALPs under the assessment of detector calibration}
\label{subsection:s6.2_marginalising_over_differentARFs}

In this section, we examine the impact of realistic imperfections in the calibration of the ARF on our ability to set bounds on the ALP-photon coupling constant using our standard $C$-statistic likelihood approach.~We simulate an ALP-free \textit{Athena}/X-IFU spectrum of NGC1275 using a particular choice of ARF.~We then fit the simulated spectrum across both our library of ALP-photon survival curves (including scans over both the ALP-photon coupling constant and the magnetic field realisations) as well as a subset of the library of calibration responses (i.e.~ARFs).~The normalised posterior distribution on ALPs must then be found by marginalising over magnetic field models and ARFs.~The Bayesian framework we follow is described in Appendix \ref{section:a2_bayesianApproach_calibration_marginalisation}.~

Marginalising over 216 ARFs and over 216 magnetic field realisations, we see that our standard $C$-stat likelihood technique for setting bounds on $g_\mathrm{a\gamma}$ is significantly influenced by imperfect knowledge of the calibration, as follows.~The excluded region weakens to $g_\mathrm{a\gamma} \geq 6.3 \times 10^{-13} \ {\mathrm{GeV}}^{-1}$ both at the $95\%$ and $99.7\% \ \mathrm{CLs}$ compared to the values of $g_\mathrm{a\gamma} > 6.3\times 10^{-14} {\mathrm{GeV}}^{-1}$ excluded at $95\% \ \mathrm{CL}$~ in the ``perfect'' calibration scenario, implying a weakening \textit{Athena}'s bounds on ALPs by an order of magnitude.

The essential reason for this is that our standard technique for assessing bounds on the ALP-photon coupling constant simply examines the significance of X-ray spectral distortions away from the astrophysical model, \textit{not} the nature of the distortions.~However, as we see from a comparison of Fig.~\ref{figure:f3_mixingCurve_example_limit-currentExclusion} and Fig.~\ref{figure:f1_effectiveArea_threeMissions}, the typical characteristics of instrumental calibration related features are very distinct from those of the ALP-photon conversion, respectively, with the latter having significant quasi-periodic structure.~In the next section, we show that the application of machine learning can use these differences to partially disentangle ALP-induced and calibration-induced effects, thereby restoring to a large degree the ability of the {\it Athena}/X-IFU to constrain ALPs.~

\section{Circumventing detector calibration limits with Neural Networks}
\label{section:s7_CNN_ML}

Because artificial neural networks (ANNs) do not rely on a fit statistic to derive a detection confidence, it is possible that they may be able to circumvent the reduction in confidence associated with calibration uncertainties.~For a proof-of-concept test we adapt the parameter estimation scheme from \citet{Parker22_agnspec_nn}, and train ANNs to recover $g_\mathrm{a\gamma}$ from raw spectra.~If a strong ALP signal is present in the spectrum, the ANN should recover the value of $g_\mathrm{a\gamma}$ of the input spectrum, and a detection limit can be derived from the point where this recovery breaks down.

Firstly, we tested several different network architectures, including the two designs from \citet[][]{Parker22_agnspec_nn}.~In the first of these architectures, the spectra were normalised and directly input into the first layer of the ANN.~In the second architecture, Principal Component Analysis \citep[][]{pearson_pca_parker22_14786440109462720} was first applied to reduce the dimensionality of the spectroscopic data (counts per energy bin), making it easier for the ANN to learn the relevant and to dismiss the redundant features of the spectroscopic data whilst reducing the risk of overfitting.~However, PCA preprocessing was found not to be effective given the complexity of the ALP-induced features in our synthetic \textit{Athena}/X-IFU spectra.
\begin{figure*}
  \centering
  \includegraphics[width=\linewidth]{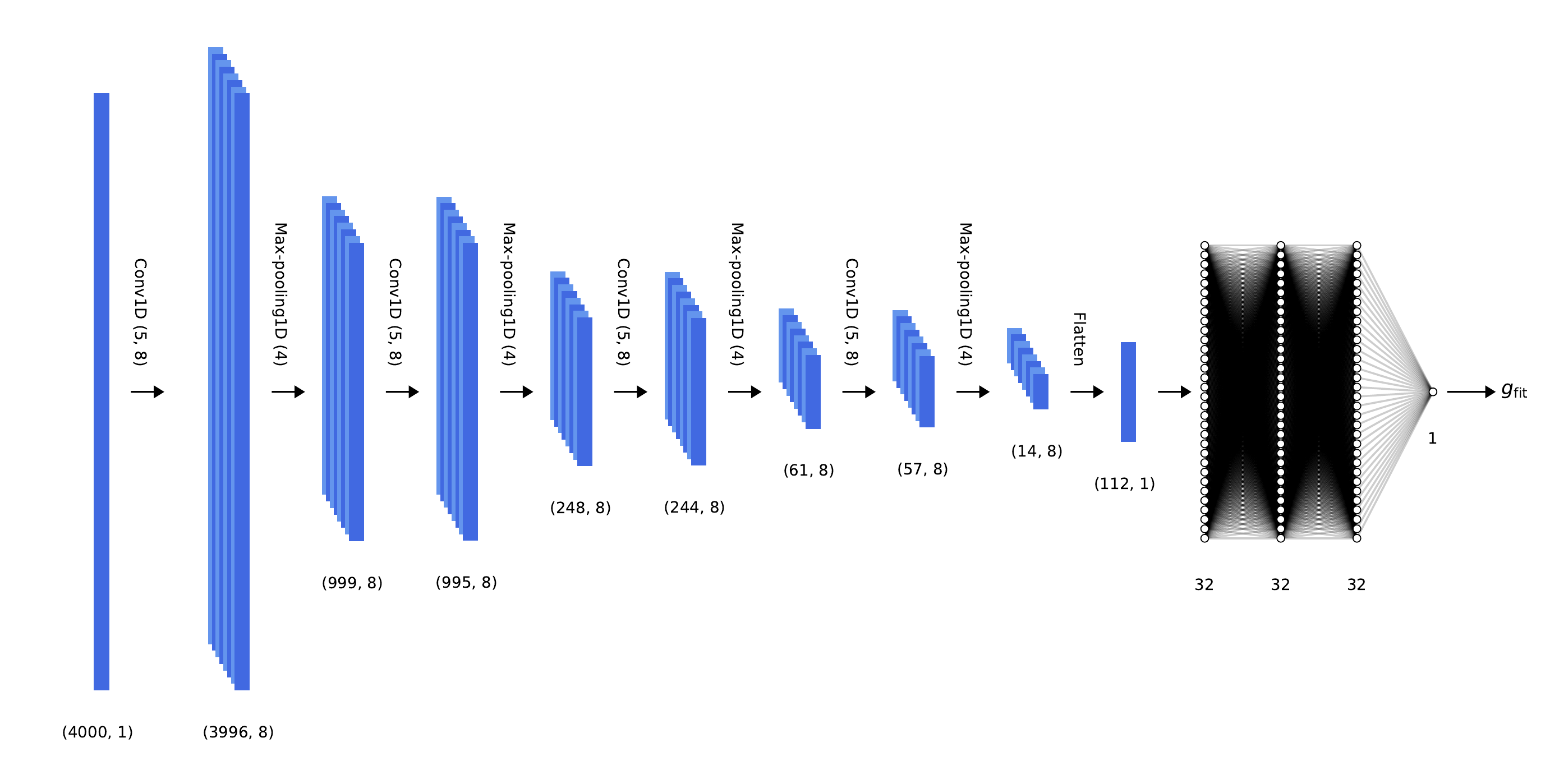}
  \caption{Architecture of the CNN used to estimate the value of $g_\mathrm{a\gamma}$.~The data input into the CNN is represented by the first blue column on the left.~This passes through a series of 1D convolution and pooling layers, which extract features from the data and then downsample it by a factor of 4 (see first paragraph of Sec.~\ref{section:s7_CNN_ML}).~The final feature map is then flattened, resulting in the final single column, before being input into a series of fully connected layers which estimate $g_\mathrm{a\gamma}$ from the feature map.~The numbers in brackets below the columns represent the size of the data or feature map at each stage in the convolutional layers, and the numbers below the fully connected layers give the number of nodes in each layer.}
  \label{figure:cnnarch}
\end{figure*}

Finally, we tested a convolutional neural network (CNN), where the spectroscopic data passes through a series of convolutional layers.~In the more common use of CNNs for image recognition, where the convolutional layers identify structure within the data, such as edges or lines, CNNs determine the important features of the data and optimise for such features themselves.~As the data passes through multiple layers, this information becomes more abstracted and more condensed.~In our CNN implementation, after each convolutional layer we include a max pooling layer, which downsamples the data, taking the highest value from each $N$ data points (where $N$ is the degree of pooling).~This compresses the data, and allows CNNs to be invariant to small shifts in the position of features.~We find that the CNN generally outperforms the two ANN architectures described above, for which we focus on the CNN for the remainder of this section.

\begin{figure*}
 \center
\includegraphics[width=0.75\textwidth]{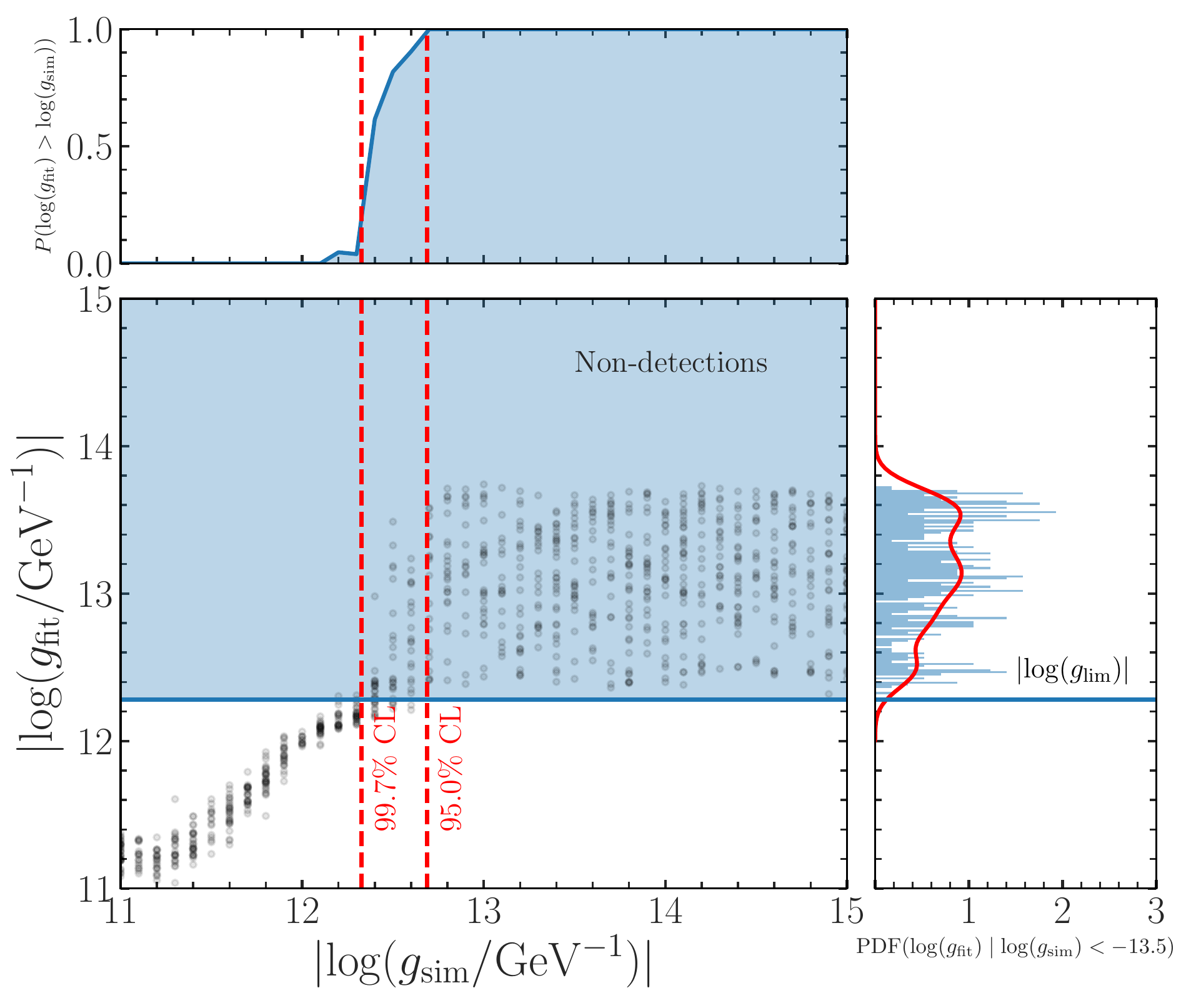}
 \caption{Main panel: recovery of the value of $g_\mathrm{a\gamma}$ as inferred by the Artifical Neural Network (ANN) introduced in Sec.~\ref{section:s7_CNN_ML}.~The x-axis of the main panel shows the absolute simulated value of the photon-ALP coupling, while the y-axis shows its parameter estimation by the ANN.~The horizontal line shows the detection threshold $|\mathrm{log}(g_\mathrm{lim})|$, defined as the 99\% limit of the distribution of $\mathrm{log}(g_\mathrm{fit})$ for $\mathrm{log}(g_\mathrm{sim})<-13.5$.~This is shown in the right panel, where the red line shows a Gaussian kernel density estimation.~Dashed lines show the limits derived from the posterior distribution of $|\mathrm{log}(g_\mathrm{sim})|$ for all spectra that do not have ALP detections (shown in the top panel).~We refer to Sec.~\ref{section:s7_CNN_ML} for discussion.}
\label{figure:recovery_nn}
\end{figure*}

The CNN was thereby trained on 10\,000 simulated \textit{Athena}/X-IFU spectra, with values of $\log(g_\mathrm{a\gamma}/{\mathrm{GeV}}^{-1}) \in [-15.0,-11.0]$ in steps of $0.1 \ \mathrm{dex}$,~and randomly allocated response files and magnetic field realisations from a library of~1\,001~representative response files and a set of 300~configurations of the turbulent cluster field.~The data are truncated to the first~20\,000~spectral bins and rebinned by a factor of~5.~We then normalised and detrended the spectra using the \textsc{scipy.signal.detrend} function.~The latter implements a linear detrending of the data.~We find our results (shown by Fig.~\ref{figure:recovery_nn}) to be insensitive to whether linear or a higher-order polynomial detrending is employed.~This applies to both the 95\% and 99.7\% exclusion levels on $g_\mathrm{a\gamma}$ inferred by the CNN and to the distributions illustrated by Fig.~\ref{figure:recovery_nn}.~We used the \textsc{Keras} deep learning API \citep{keras} for \textsc{python}, running on the \textsc{TensorFlow} machine learning platform \citep[][]{tensorflow}.~We used the \texttt{Adam} optimiser \citep{Kingma14}, the Rectified Linear Unit (ReLU) activation function \citep{Nair2010rectified}, the mean squared error loss function, and early stopping \citep{Prechelt1998early} with a patience of 50.~We followed each convolutional layer with a MaxPooling layer, downsampling the feature map.~Our CNN design is summarised in Fig.~\ref{figure:cnnarch}.

We performed a basic hyperparameter optimisation, randomly selecting 200 combinations of values of the number of convolutional layers, number and breadth of dense layers, number of filters, learning rate, batch size, and pooling size.~In each case, we evaluated the performance of the network on a test set of~2\,000~spectra and selected the best performing network based on the mean-squared error (Fig.~\ref{figure:cnnarch}).~Finally, we used the network to predict the value of $g_\mathrm{a\gamma}$ from a set of 1\,001 validation spectra (one for each response file) which were not used to train the network or evaluate the performance during hyperparameter optimisation.~We note that the latter were generated with an independent library of 204 magnetic field configurations within Perseus.

The recovered value of the photon-ALP coupling as a function of its simulated value is shown by the main panel in Fig.~\ref{figure:recovery_nn}.~We refer to these variables as $\log(g_\mathrm{sim})$ and $\log(g_\mathrm{fit})$, respectively, where both $g_\mathrm{sim}$ and $g_\mathrm{fit}$ are expressed in $\mathrm{GeV}^{-1}$.~The absolute value of these variables is shown in Fig.~\ref{figure:recovery_nn},~where the photon-ALP coupling value decreases along both axes.

We interpret the CNN regression method by referring the reader to the main panel of Fig.~\ref{figure:recovery_nn}.~For sufficiently high values of $g_\mathrm{a\gamma}$, the photon-ALP induced oscillations are distinguishable from the effects of detector calibration.~Clearly, the CNN is successful at identifying the presence of ALPs when $\mathrm{log}(g_\mathrm{sim}) \gtrsim -12.3$, albeit with a bias of 0.2-0.4 dex in $\mathrm{log}(g_\mathrm{rec})$.~The flattening of $g_\mathrm{fit}$ at $|\mathrm{log}(g_\mathrm{sim})| \in [11.0,11.2)$ arises due to the averaging of $P_\mathrm{a\gamma}$ to 0.5 at hard energies from an unpolarized photon beam~\citep[see~Sec.~4 of][]{Marsh17}.

In the intermediate range of $\mathrm{log}(g_\mathrm{sim}) \in [-12.7, -12.4]$, the CNN is only partially successful at detecting the presence of ALPs in the input spectra.~In such successful scenarios, one would expect a favourable magnetic field model that enables the CNN to correctly disentangle between the effects of: detector calibration, ALP-induced features and Poisson noise in the synthetic data.~However, for simulated values of the photon-ALP coupling below the intermediate range,~i.e.~for $\mathrm{log}(g_\mathrm{sim}) \leq -12.8$, the CNN is unable to recover the ALP signal as the amplitude and shape of the ALP-induced perturbations are comparable and indistinguishable to those of the Poisson noise in the data.~In this regime, the CNN ``guesses'' a value of $g_\mathrm{fit}$ within the range of $g_\mathrm{sim}$ which the CNN cannot adequately recover in order to minimise the loss function.

We compute the 95\% and 99.7\% confidence levels (CLs) on $g_\mathrm{a\gamma}$ inferred by the CNN as follows.~To begin with, we consider the probability of obtaining a given value of $\mathrm{log}(g_\mathrm{fit})$ from $\mathrm{log}(g_\mathrm{sim})$.~First, we define a detection threshold.~We use the points $|\mathrm{log}(g_\mathrm{sim})| \geq 13.5$ as proxies for spectra with no injected ALP signal.~We calculate the distribution of~$\mathrm{log}(g_\mathrm{fit})$ for these spectra.~The latter is shown by the right panel of Fig.~\ref{figure:recovery_nn}.~We then define a detection (``limit'') threshold, $\mathrm{log}(g_\mathrm{lim})$, such that there is a 1\% false positive rate.~We then find the posterior distribution of $\mathrm{log}(g_\mathrm{sim})$ for all spectra that have $\mathrm{log}(g_\mathrm{fit})>\mathrm{log}(g_\mathrm{lim})$.~The probability this distribution corresponds to is shown by the top panel of Fig.~\ref{figure:recovery_nn}.~Adopting the same priors in ALP parameters described in Sec.~\ref{section:s5_nominal_nextGen}, we extrapolate this out to $|\mathrm{log}(g_\mathrm{a\gamma}/{\mathrm{GeV}}^{-1})|=19.0$, and then find the 95\% and 99.7\% CLs of the distribution.

In general, one would expect the CNN performance to readily improve upon training it with a larger sample of synthetic spectra and with the use of alternative network architectures.~This will be the subject of a future publication.~We highlight that training the CNN on residuals from fitting synthetic data -rather than on spectra- could also help improve the CNN's performance \citep[see][]{Schallmoser2021_updated_ML}.

Clearly, the CNN permits inferring tighter projected bounds on ALPs from \textit{Athena}/X-IFU under the assessment of detector calibration when compared to the $C$-stat likelihood approach (Sec.~\ref{subsection:s6.2_marginalising_over_differentARFs}).~We refer to Sec.~\ref{section:ml_discussions} for further discussion and for an overview of other ML implementations in the context of cluster/ALP searches.~We highlight that ML classifiers such as Support Vector Machines, Decision Tree and Random Forest classifiers were used by \citet[][]{Schallmoser2021_updated_ML} to tighten the bounds on ALPs previously found by \citet[][]{Conlon_17_manySources}.


\section{Injected signal: \textit{Athena} recovery}
\label{section:s7_injSignal}
 \begin{figure*}
 \center
 \includegraphics[width=1.\textwidth]{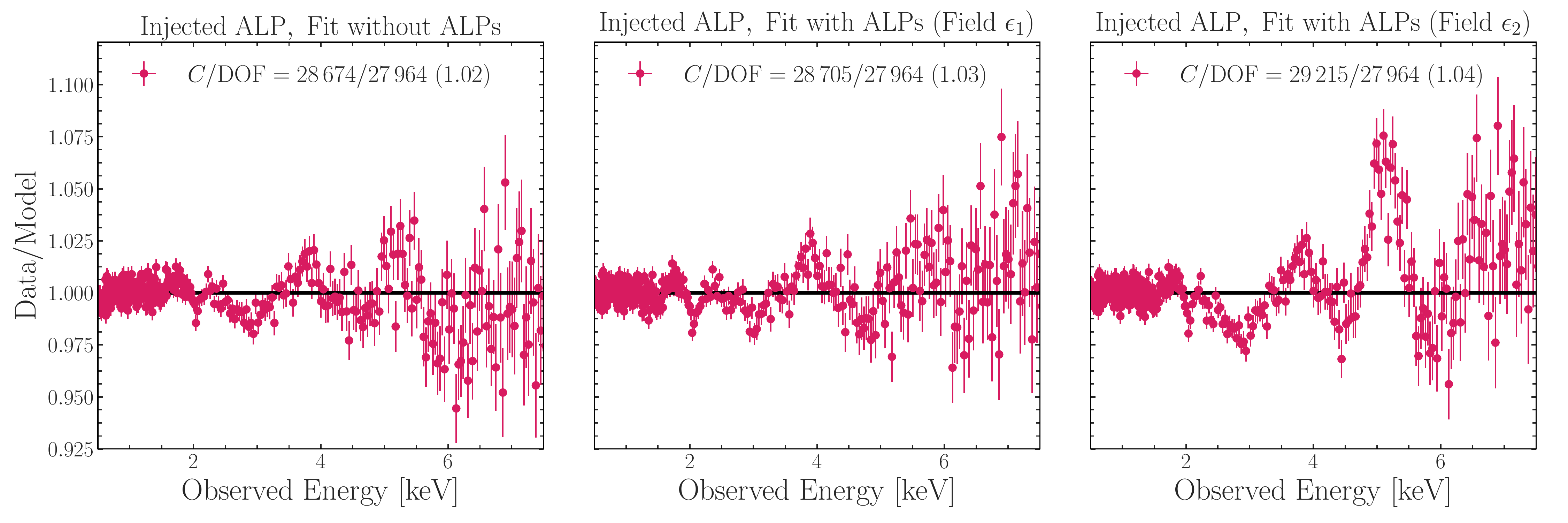}
 \caption{Residuals from fitting one of the \textit{Athena}/X-IFU injected ALP spectra presented in Sec.~\ref{section:s7_injSignal} with a set of spectral models.~For illustration purposes, all panels show the residuals under the optimal calibration scenario.~The photon-ALP model employed as the injected signal is shown in the bottom panel of Fig.~\ref{figure:f1_effectiveArea_threeMissions}, and features a ``true'' value of $g_\mathrm{a\gamma} = 5.0\times10^{-13} \ {\mathrm{GeV}}^{-1}$.~The different panels illustrate the importance of the field model when employing a cell-based approach to describe the spectrum, as follows.~Left panel: a ``fiducial'' (astrophysics-only) model is fitted.~Central and right panels: a multiplicative ALP model (featuring $g_\mathrm{a\gamma} = 5.0\times 10^{-13} \ {\mathrm{GeV}}^{-1}$) is fitted under a specific turbulent field geometry (referred to as $\epsilon_1,\epsilon_2$).~The simulated data was fitted within the observed energies $0.5 - 12 \ \mathrm{keV}$.~For plotting purposes, a binning scheme of a target signal-to-noise ratio 250 with the restriction that no more than 100 spectral bins were coadded was applied, and we show the energy range $0.5 - 7.5 \ \mathrm{keV}$.~The best-fit $C$-stat of each fit, the number of degrees of freedom (DOF), and the reduced $C$-stat ($\equiv C/\mathrm{DOF}$) are quoted (with the latter in parenthesis).}
 \label{figure:f_injSignal_residuals}
\end{figure*} 

\noindent A useful exercise that facilitates comparison with Sec.~\ref{section:s7_CNN_ML} within the $C$-stat likelihood framework we have employed (Secs.~\ref{section:s5_nominal_nextGen} and \ref{subsection:s6.2_marginalising_over_differentARFs}) is to assess to which extent \textit{Athena} would be able to recover a true ALP signal under the effects of detector calibration, as follows.

We firstly simulated an \textit{Athena}/X-IFU $200$-$\mathrm{ks}$ observation of NGC1275 by loading a multiplicative ALP spectrum with $m_\mathrm{a} = 10^{-14} \ \mathrm{eV}$ and $g_\mathrm{a\gamma} = 5.0\times 10^{-13} \ {\mathrm{GeV}}^{-1}$ (whose photon survival curve is shown in the lower panel of Fig.~\ref{figure:f3_mixingCurve_example_limit-currentExclusion}).~As in Sec.~\ref{section:s3_simulated_spectra_ngc1275_nominal}, we considered Poisson noise statistics without the assignment of a background spectral file.~This “injected” spectrum was generated under a turbulent cluster field configuration which is not part of the 216 set of models present in our analysis pipeline~(Sec.~\ref{section:s4_alpGrid}).

The effects of the turbulent magnetic field model in the recovery of the ALP signal are shown by Fig.~\ref{figure:f_injSignal_residuals}.~Its left panel shows the residuals resulting from fitting the injected spectrum with an astrophysics-only model under a given detector calibration setup.~Clearly, the injected ALP induces features in the spectrum such that residuals up to~$\sim 2\%$~appear in the spectrum at high energies.

The central and right panels of Fig.~\ref{figure:f_injSignal_residuals} illustrate the residuals resulting from fitting the injected spectrum with two multiplicative photon-ALP mixing models ($\epsilon_1$ and $\epsilon_2$, see upper panel of Fig.~\ref{figure:f3_mixingCurve_example_limit-currentExclusion}) with $m_\mathrm{a} = 10^{-14} \ \mathrm{eV}$ and $g_\mathrm{a\gamma} = 5.0\times 10^{-13} \ {\mathrm{GeV}}^{-1}$.~Interestingly, the geometry of field $\epsilon_2$~(Fig.~\ref{figure:f_injSignal_residuals}) is better suited to describing the injected ALP spectrum at energies $4 - 6 \ \mathrm{keV}$~(Fig.~\ref{figure:f_injSignal_residuals}).

The injected spectrum was thereby fitted across our library of 216 possible detector responses, each of which was fitted across our library of 216 field models, following our previous analysis.~The posteriors on ALPs were inferred following the Bayesian framework described in Appendix \ref{section:a2_bayesianApproach_calibration_marginalisation}.~The injected signal was found to be well recovered at both the $95\%$ and $99.7\% \ \mathrm{CLs}$, that is, all couplings $g_\mathrm{a\gamma} \geq 5.0\times {10}^{-13} {\mathrm{GeV}}^{-1}$ are excluded at the $2\sigma$ level for $m_\mathrm{a} < 10^{-12} \ \mathrm{eV}$.~

\begin{figure}
 \center
 \includegraphics[width=0.45\textwidth]{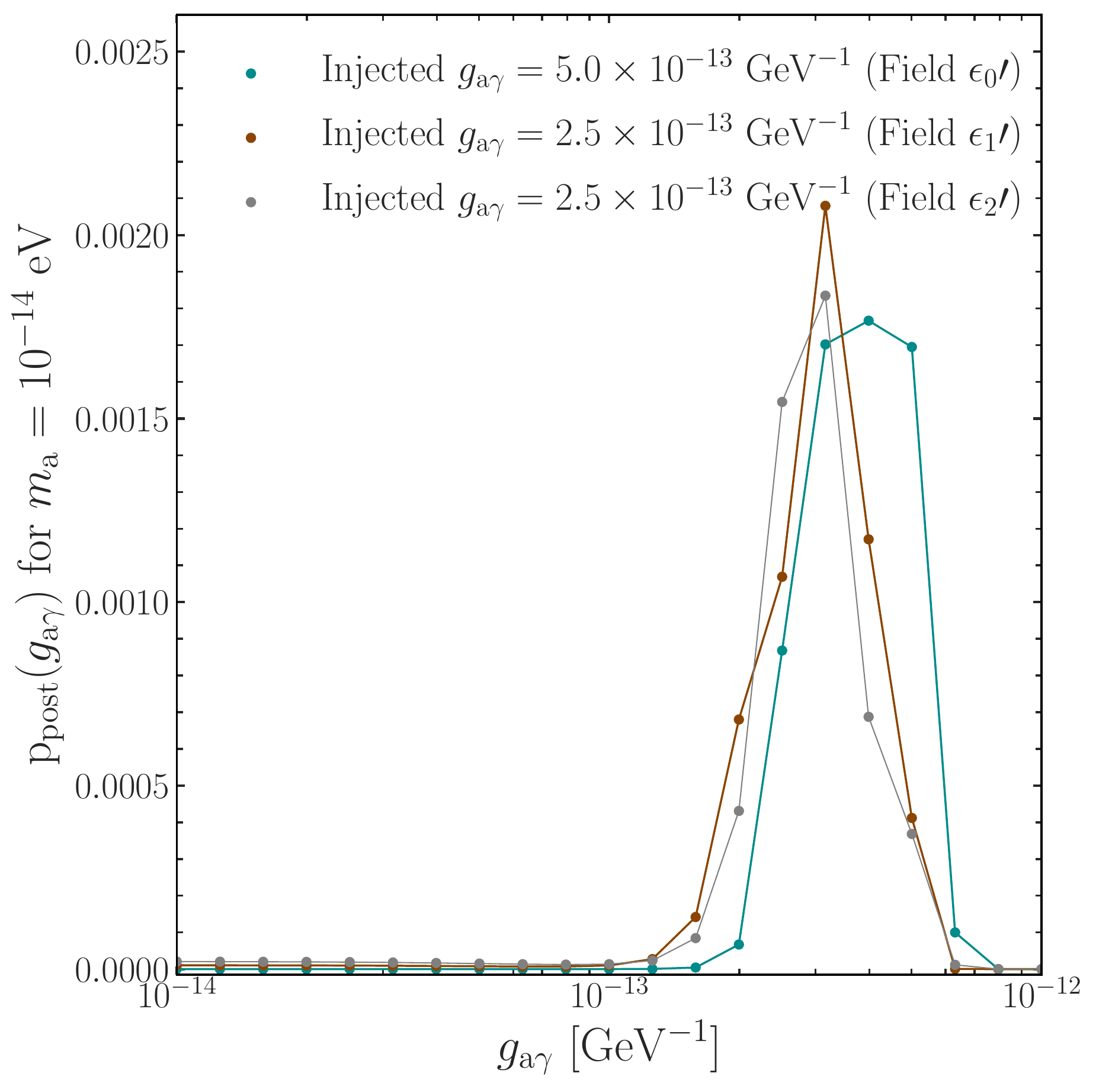}
 \caption{Normalised posterior distribution on $g_\mathrm{a\gamma}$ for ALPs of $m_\mathrm{a} = 10^{-14} \ \mathrm{eV}$ (see Eq.~\ref{eq:eq6_normalisation_condition}).~Each colour represents the posterior inferred when fitting each of the simulated \textit{Athena}/X-IFU spectra of NGC1275 which contain a specific ALP signal (illustrated in the lower panel of Fig.~\ref{figure:f_injSignal_residuals}).~In all cases, the posteriors were inferred by marginalising over 216 magnetic field realisations and over 216 conservative detector responses.}
 \label{figure:f7_posteriors_injected}
\end{figure}

We note that, given the input spectrum contains an ALP in the ``massless regime'', no knowledge on the ALP mass would be recovered and would be limited to $\mathrm{log}(m_\mathrm{a}/\mathrm{eV})\in [-30.0,-12.0)$.~Nevertheless, \textit{if} there was an effectively massless ALP with $g_\mathrm{a\gamma} = 5.0\times 10^{-13} \ {\mathrm{GeV}}^{-1}$ in the spectrum, it would indeed be detected by \textit{Athena}/X-IFU, even under the assessment of detector calibration.~Fig.~\ref{figure:f7_posteriors_injected} shows the posterior on the recovery of such an injected signal under the effects of detector calibration, where the inferred posterior on $g_\mathrm{a\gamma}$ is Gaussian-distributed with a mean corresponding to the true $g_\mathrm{a\gamma}$~and a standard deviation of $0.1 - 0.2 \ \mathrm{dex}$.

On the other hand, the upper bound on $g_\mathrm{a\gamma}$ inferred in the previous section (see Fig.~\ref{figure:recovery_nn}) suggests that one could expect to recover all true $g_\mathrm{a\gamma}$ up to $\mathrm{log}(g_\mathrm{a\gamma}/{\mathrm{GeV}}^{-1}) \sim-12.6$.~We thereby generated two simulated \textit{Athena}/X-IFU spectra, with $200$-$\mathrm{ks}$ of exposure, by loading multiplicative ALP models with $m_\mathrm{a} = 10^{-14} \ \mathrm{eV}$ and $\mathrm{log}(g_\mathrm{a\gamma}/{\mathrm{GeV}}^{-1}) =-12.6$ using two configurations of the cluster field not embedded in the photon-ALP mixing grid introduced in Sec.~\ref{section:s4_alpGrid}.~Their corresponding photon survival functions are shown in the lower panel of Fig.~\ref{figure:f3_mixingCurve_example_limit-currentExclusion}.~Both of these spectra were then fed into our analysis scripts, resulting in the posterior distributions shown in Fig.~\ref{figure:f7_posteriors_injected}.~Although biased to higher photon-ALP coupling values, both signals would be detected and recovered.~This suggests that, even if marginally outside of the detection threshold, a true ALP signal may eventually be recovered if existing in a favourable underlying cluster field geometry.~We refer to Sec.~\ref{section:sub9.25_fieldModel} for further discussion.~

\section{Discussion}
\label{section:s9_discussions}
\subsection{Bounds on ALPs from Athena and AXIS: optimal calibration scenarioes}
\label{section:sub9.1_summary_results_calibration_athena} \noindent Following from the work of \citet[][]{Conlon_2017_ProjectedBoundsOnAthena}, we have shown that \textit{Athena}/X-IFU, with its unprecedented effective area and spectral resolution, can in principle improve on the current ALP constraints (Tab.~\ref{table:t0_alpBounds_current_andNominal_nextGen}) by an order of magnitude for $\mathrm{log}(m_\mathrm{a}/\mathrm{eV})<-12.0$, \textit{if} its calibration is well understood.

In this work, we simulated a $200$-$\mathrm{ks}$ \textit{Athena}/X-IFU observation of NGC1275.~The cluster emission component which would be included in the {\it Athena} beam was modelled with a Doppler-broadened single-plasma temperature component, with reference free parameters taken from \citet[][]{atomicdata_hitomi_2018} (see Sec.~\ref{section:sub9.4_systematicsAndLimitations} for further details).~We highlight that, for a $1$-$\mathrm{Ms}$ \textit{Athena}/X-IFU observation of NGC1275, all photon-ALP couplings $g_\mathrm{a\gamma} > 4.0\times 10^{-14} \ {\mathrm{GeV}}^{-1}$ would be excluded for $m_\mathrm{a} \leq 10^{-12} \ \mathrm{eV}$ under an optimal calibration scenario.~Following from \citet[][]{Reynolds20}, we have employed a fiducial model for the cluster magnetic field (summarised in Sec.~\ref{section:sub9.25_fieldModel}).~In our analysis pipeline, we consider 216 realisations of the turbulent cluster field.
 
We have also presented projected bounds on effectively massless ALPs from \textit{AXIS}, under its current target and baseline on-axis response files.~In both cases, we generated a fake $200$-$\mathrm{ks}$ \textit{AXIS} observation of NGC1275, finding an upper bound of $g_\mathrm{a\gamma} \sim 2.0 \times 10^{-13} \ {\mathrm{GeV}}^{-1}$ at the $95\%$ level for both responses.~With its more modest spectral resolution compared with that expected for \textit{Athena}/X-IFU, \textit{AXIS}’s superb spatial resolution will permit a clear extraction of the intrinsic AGN emission.~This will provide \textit{AXIS} with exciting prospects to constrain ALPs, in particular, given the improvement compared to the current most sensitive constraints from single-source \textit{Chandra} AGN observations (Tab.~\ref{table:t0_alpBounds_current_andNominal_nextGen}).~

Our projected \textit{AXIS} bounds on ALPs also highlight the prospects of next-generation telescopes with intrinsically different setups to probing the physics Beyond the Standard Model such as \textit{Arcus}, \textit{LEM} \citep[][]{siskreynes2023current},~\textit{STROBE-X} \citep[][]{strobex} and \textit{Lynx} \citep[][]{lynxwhitepaper}.
\subsection{Bounds on ALPs from Athena: conservative calibration scenario}

Knowledge of detector calibration will undoubtedly make an impact on the potential of next-generation observatories to constraining Physics Beyond the SM and, in particular, very-light ALPs.~In our work, we have solely focussed on reassessing \textit{Athena}'s projected bounds under the effects of calibration, although such a consideration will need be made for any upcoming X-ray missions.~

We first employed a library of conservative calibration files and fitted the simulated $200$-$\mathrm{ks}$ \textit{Athena}/X-IFU observation of NGC1275 across such library, and 216 models for the turbulent field for each detector response.~Using the standard $C$-stat likelihood procedure (see Appendix \ref{section:a2_bayesianApproach_calibration_marginalisation}), marginalising over field models and detector responses, all couplings $g_\mathrm{a\gamma} > 6.3 \times 10^{-13} \ {\mathrm{GeV}}^{-1}$ for ALPs of masses $m_\mathrm{a} \leq 10^{-12} \ \mathrm{eV}$ are excluded at the $95\%$ and $99.7\%$ levels.~This would imply a weakening of \textit{Athena}'s projected bounds on massless ALPs by an order of magnitude, compared to the optimal calibration scenario.~We note that, for a $1$-$\mathrm{Ms}$ \textit{Athena}/X-IFU observation of NGC1275, all photon-ALP couplings $g_\mathrm{a\gamma} > 4.0 \times {10}^{-13} \ {\mathrm{GeV}}^{-1}$ would be excluded at the $95\%$ CL on the basis of the standard $C$-stat likelihood procedure.

With the aim of disentangling between the effects of cross-calibration residuals and those of photon-ALP mixing in order to acquire more sensitive bounds on ALPs, we proceeded by reevaluating \textit{Athena}'s projected bounds on ALPs through the use of Machine Learning.

\subsection{The effect of calibration assessed via Machine Learning}
\label{section:ml_discussions}
Machine Learning (ML) techniques have increasingly been used in cluster-ALP searches \citep[e.g.][]{Conlon_Rummel_2019_stat_withML, franDay_20_acceleratingWithML}.~One of the most relevant studies pertinent to our work is that of \citet[][]{Schallmoser2021_updated_ML}, which employs ML classifiers to find updated bounds on very-light ALPs from previous \textit{Chandra} observations of AGN centred or behind clusters, following from the work of \citet[][]{Conlon_17_manySources}.~Under a 3-dimensional~field model (see Sec.~\ref{section:sub9.25_fieldModel} for further discussion) and for the multiclass classification method, \citet[][]{Schallmoser2021_updated_ML} were able to tighten the previous upper bound on $g_\mathrm{a\gamma}$ from \textit{Chandra} observations of A1795Sy1 by a factor of 4, excluding $g_\mathrm{a\gamma} \gtrsim 6 \times 10^{-13} \ {\mathrm{GeV}}^{-1}$ at the $95\%$ level (Tab.~\ref{table:t0_alpBounds_current_andNominal_nextGen}).~In this case, the ML classifier had been trained on the residuals from fitting simulated \textit{Chandra} spectra of the source divided by the average survival probability across a range of $g_\mathrm{a\gamma}$ values whilst being fitted by an astrophysics-only model.~Finally, we note that \citet[][]{Schallmoser2021_updated_ML} were the first to use Approximate Bayesian Computation Methods (ApBC), a specific type of likelihood-free inference, to find updated bounds on $g_\mathrm{a\gamma}$ from their analysis \citep[][]{Sisson_overview_ApBC}.~

Our ML study mostly differs from that of \citet[][]{Schallmoser2021_updated_ML} as follows.~\citet[][]{Schallmoser2021_updated_ML} use a wide range of machine-learning classifiers to distinguish between spectra with and without an injected ALP signal.~The classifiers had been trained on spectra with a specific coupling, and were then tested on spectra with a range of couplings.~We instead test a convolutional neural network (CNN) regression approach, training CNNs on data with a wide range of coupling factors to estimate the coupling as accurately as possible.~Our parameter estimation approach more closely mirrors conventional spectral fitting.~

The advantage of a convolutional approach is that it enables the network to effectively ``learn'' the important features of the data for itself, for example by detecting edges in the data, and also makes it invariant to small shifts of spectral features due to the pooling of features from adjacent bins.~We speculate that the convolutional layers are able to distinguish between the shapes of residual features caused by any instrumental miscalibration and residual features due to an ALP signal based on their shapes.~That is, ARF features tend to be either sharp edges or gradual broadband slopes, whereas ALP features are smoother (that is,~more absorption-like).~For instance, the amplitude of the ALP signal tends to increase monotonically with energy (Fig.~\ref{figure:f3_mixingCurve_example_limit-currentExclusion}).

The biggest limitation of the machine-learning approach is likely the requirement for synthetic training data.~This means that the accuracy and performance of the CNN is dependent on how comprehensive the simulated spectra are, as the network cannot be relied upon outside the scope of the training set.~The ANNs used here are relatively simple compared to many used in deep learning applications, so they can be trained and tested in a short timeframe ($\sim5$~minutes running on a laptop CPU) once the synthetic training set has been constructed.

\subsection{Modelling Cluster Magnetic fields}
\label{section:sub9.25_fieldModel}

Our inferred upper bounds on $g_\mathrm{a\gamma}$ are based upon the assumption that the magnetic field strength is scaled by the electron number density for an isothermal ICM.~The inferred field profile broadly reproduces a value of the magnetic-to-pressure ratio of $\beta_\mathrm{plasma} = 100$ up to $1.8 \ \mathrm{Mpc}$ up from the cluster centre.~This fiducial field model, which had already been adopted in `Model B' of \citet[][]{Reynolds20}, is supported by RM observations of cool-core clusters \citep[][]{Taylor06, govoni_2012_summary_bfields_icm} and by measurements of the turbulent velocities of the ICM in the Perseus cluster \citep[][]{hitomi_gasvel_turb_18, Zhuravleva2014_turbulentPerseusVirgo}.~Moreover, Magnetohydrodynamic (MHD) simulations of galaxy clusters predict the existence of turbulent magnetic fields in the ICM \citep[][]{Donnert2018_MagneticFieldReview, amplificationFields_vazza}.~However, some of these studies have predicted the radial decrease of $\beta_\mathrm{pl}$ across the cluster volume, e.g.~due to the increase in non-thermal pressure support triggered by mergers.~We refer to \citet[][]{2022_matthews_bfieldModels} and \citet[][]{Marsh2021_Fourier} for discussion of the impact of magnetic field modelling assumptions on photon-ALP bounds inferred from cluster/AGN searches under cell-based and Gaussian Random Field (GRF) approaches.~Specifically, \citet[][]{2022_matthews_bfieldModels} found that a radially-dependent $\beta_\mathrm{pl}$ can systematically shift bounds on $\mathrm{log}(g_\mathrm{a\gamma}/{\mathrm{GeV}}^{-1})$ from cell-based and GRF studies by $0.3 \ \mathrm{dex}$ (depending on the exact radial profile of $\beta_\mathrm{pl}$).~The same study also finds that the limits are fairly insensitive to whether cell-based or Gaussian Random Field (GRF) models are used.~

The use of the Fourier formalism \citep[][]{Marsh2021_Fourier}, accurate to leading order in the coupling, may provide a fairly computationally inexpensive route to computing photon-ALP conversion.~This formalism is valid for sufficiently low values of $g_\mathrm{a\gamma}$ such that the ALP-induced distortions are of the order~$\lesssim 5-10\%$,~which is the appropriate regime for our simulated observations.~This formalism, embedded in \texttt{ALPRO} \citep[][]{alpro_james_2022_zenodo}, can also be applied to the massive ALP regime \citep[see Sec.~III of][]{Marsh2021_Fourier}.~Furthermore, as noted by \citet[][]{Marsh2021_Fourier}, if an ALP signal is present, the residuals of the data may be able to be transformed directly to obtain the autocorrelation function of the line-of-sight magnetic field.~Such an exercise is challenging, requiring very good quality data, but the high spectral resolution and large effective area of {\sl Athena} offer exciting prospects if this formalism were to be applied in future.~

Ideally, a 3-dimensional description of the cluster field, constrained by observational RM and pressure profile data, would be adopted to unveil its true underlying structure \citep[for recent studies, see][]{Schallmoser2021_updated_ML, pierluca_mhd_2022}.~The turbulent nature of the field can create departures from predictions using a GRF, as shown by \cite{pierluca_mhd_2022}.~They compute the conversion probabilities using a high-resolution 3D MHD simulation, finding that non-Gaussian, local spikes in the MHD magnetic field produce `heavy-tailed' distributions of the conversion probability.~This work therefore suggests that, for certain sightlines, stronger ALP signals could be observed.

Finally, we note that \textit{if} an effectively massless ($m_\mathrm{a} \lesssim 10^{-12} \ \mathrm{eV}$) ALP signal with sufficiently high $g_\mathrm{a\gamma}$ (compared to the Poisson noise) were to be discovered by \textit{Athena}/X-IFU, the extent to which the former signal would be recovered would be depend on the underlying field structure.~This is shown by Figs.~\ref{figure:f_injSignal_residuals} and \ref{figure:f7_posteriors_injected}; the results are encouraging, in the sense that the ALP coupling is recovered quite reliably even when the exact magnetic field structure is not known.~However, in this effectively massless ALP regime, no information about the true ALP mass would be retained, given that all ALPs of masses $\mathrm{log}(m_\mathrm{a}/\mathrm{eV}) \in [-30.0, -12.0)$ induce near-identical spectral distortions.~In addition, if the true magnetic field model intersected by the line-of-sight included the non-Gaussian structure described by \cite{pierluca_mhd_2022}, then using a cell-based or GRF model for the inference of $g_\mathrm{a\gamma}$ would result in a bias, leading to a higher $g_\mathrm{a\gamma}$ being estimated than the true photon-ALP coupling value.

Overall, the above considerations regarding magnetic field modelling suggest a fairly encouraging outlook for the next generation of X-ray ALP searches, but they do highlight the importance of high frequency-resolution Rotation Measure studies of clusters, combined with improved theoretical understanding of the ICM magnetic field structure from the cluster core to the virial radius.~

\subsection{Other systematics}
\label{section:sub9.4_systematicsAndLimitations} 
\subsubsection{Initially unpolarised photon beam}
In our work, we have solved the photon-ALP propagation problem for an initially unpolarised photon beam propagating through the host cluster.~One would expect such an assumption to yield conservative photon-ALP mixing probabilities compared to an initially polarised photon beam.~We refer the reader to \citet[][]{Day_2018_polarimeters_alps, galanti_polarisation_overview, galanti22_clusterPolarisation, Dessert_2022_MWD_polarisation} for explorations on how one can use source polarisation and ALP-induced polarisarion to probe ALPs.

\subsubsection{Spectral model for NGC1275}
Other than for its normalisation, all model parameters of the \texttt{bvvapec} model component applied to describe the simulated \textit{Athena}/X-IFU spectrum of NGC1275 were frozen (see Tab.~\ref{table:t2_astroModel_specifications_baseline}).~Our inferred posterior on ALPs remain insensitive to this choice.~Furthermore, one would expect our results to remain insensitive to whether partial covering is accounted for in the fits to our \textit{Athena} and \textit{AXIS} simulated spectra of NGC1275 \citep[see Fig.~7 and discussion in][]{2022_matthews_bfieldModels}.

Finally, we note that throughout our work, we ignored the effects of relativistic reflection from slowly-moving material surrounding the immediate vicinity of the NGC1275.~The latter has frequently been modelled to explain the $6.4$-$\mathrm{keV}$ reflection signature frequently detected in high-resolution X-ray spectra of NGC1275 \citep[e.g.][]{2018_hitomi_resolved_fekalpha_line,2021_reynolds_circumnuclearEnv_ngc1275}.~We would expect this to not affect the photon-ALP bounds inferred in our work.

\section{Conclusions}
\label{section:s10_conclusions}

We have conducted an analysis that revisits \textit{Athena}'s projected bounds on very-light ALPs, following from \citet[][]{Conlon_2017_ProjectedBoundsOnAthena}, under the effects of detector calibration.~We have also explored the potential of \textit{AXIS} for Beyond the Standard Model searches.~In both cases, we employ $200$-$\mathrm{ks}$ simulated observations of NGC1275 as the subject of our study; for concreteness, we consider detector setups broadly consistent with those proposed in \citet[][]{Barret2020_AthenaDescription} and \citet[][]{2019_axismission}, respectively.~Our main findings are:
\begin{enumerate} 
  \item Bright cluster-hosted AGN located within magnetically rich galaxy clusters are excellent probes of massless ALPs, $\mathrm{log}(m_\mathrm{a}/\mathrm{eV})<-12.0$.~With its unprecedented spectral energy resolution ($2.5 \ \mathrm{eV}$ for its X-IFU), \textit{Athena} will be able to exclude all photon-ALP couplings $g_\mathrm{a\gamma} \geq 2.0 \times 10^{-13} \ {\mathrm{GeV}}^{-1}$ at $95\%$ confidence level (CL).~These bounds have been inferred under a conservative assessment of detector calibration.~
  \item \textit{Athena}/X-IFU will be able to improve on the current limits on $g_\mathrm{a\gamma}$ by an order of magnitude (excluding $g_\mathrm{a\gamma}\geq 6.3\times 10^{-14}\,{\rm GeV}^{-1}$ at 95\% confidence level) provided an accurate knowledge of its detector-induced residuals.~For a $1$-$\mathrm{Ms}$ exposure of NGC1275, the upper bound on $\mathrm{log}(g_\mathrm{a\gamma}/{\mathrm{GeV}}^{-1})$ would further tighten by $0.2~\mathrm{dex}$~in both optimal and conservative calibration scenarioes.
  \item Machine Learning will provide an automated route to disentangling the effects of detector calibration and photon-ALP induced residuals in next-generation cluster/ALP searches.~This will be particularly relevant for the \textit{AXIS} CCD mission.~
  \item Despite having a fundamentally different design to \textit{Athena}/X-IFU, \textit{AXIS}'s more moderate spectral resolution will be circumvented by its unprecendented angular resolution, being able to offer high-quality AGN spectra free from cluster emission.~In the optimal calibration scenario, \textit{AXIS} would exceed the current best bounds on light ALPs from a $200$-$\mathrm{ks}$ on-axis observation of NCG1275 by a factor of 3.~Excitingly, for a $1$-$\mathrm{Ms}$ observation of NGC1275, \textit{AXIS} will exclude all couplings $g_\mathrm{a\gamma} > 4\times 10^{-14} \ {\mathrm{GeV}}^{-1}$ in the optimal calibration scenario, exceeding the current best bounds on massless ALPs by more than an order of magnitude.
  \item Next-generation X-ray observatories may play a key role in constraining axions arising in string theories which predict low values of $g_\mathrm{a\gamma}$ such as type-IIB string theories \citep[][]{demirtas_20_axiverse, Halverson19StringTheories}.~Some of these theories will additionally make predictions for a favoured number of ultra-light axion fields.~The parameter space of these theories may additionally be probed by complementary astrophysical observations, e.g.~with population studies of the spin/mass distribution of black holes over cosmic time with next-generation gravitational wave detectors such as \textit{LISA} \citep[][]{djemarsh_calabiYau_stringTheory_superradiance}.
\end{enumerate}
Overall, our work shows the potential of next-generation X-ray observatories to constraining very-light ALPs, even under fundamentally different designs.~Our analysis suggests that, in future, these X-ray missions may complement future laboratory-based ALP Dark Matter searches (e.g.~in birefringent cavities such as \textit{ADBC}) down to very-light masses.~

\section*{Data Availability}
The simulated X-ray spectra of NGC1275 which this study is based on were generated from the optimal Response Matrix and Area Response Files of the \textit{Athena} \citep[][]{athena_nominal_calibration} and \textit{AXIS}\footnote{See \hyperlink{}{https://axis.astro.umd.edu/}} telescopes.~The non-optimal ARFs over which we marginalise (Sec.~\ref{subsection:s6.2_marginalising_over_differentARFs}) were kindly provided by D.~Barret.~Moreover, the raw X-ray data employed to estimate the normalisation factor of the single-temperature cluster emission component (Sec.~\ref{subsubsection:axis_sim_spectra_nominal}) in our simulated \textit{Athena}/X-IFU spectrum of NGC1275 is available from the public data archives of the Chandra Science Center.~Our limits data will be made publicly available in the \texttt{ALPRO}\footnote{See \hyperlink{}{{https://github.com/jhmatthews/alpro/tree/v1.1}{}}} repository.~The reduced data products used in this work may be shared on reasonable request to the authors.

\section*{Acknowledgements}
We thank the anonymous referee for a constructive and helpful report.~We also thank Didier Barret for providing access to the library of {\it Athena}/X-IFU response files from \citet[][]{Barret_and_cappi_2019_bhspin_degen} as well as for providing useful comments on this paper.~We thank Francesca Chadha-Day, Andy Fabian and Stefan Heimersheim for helpful discussions.~J.S.R acknowledges the support from the Science and Technology Facilities Council (STFC) under grant ST/V50659X/1~(project reference 2442592).~C.S.R.~thanks the STFC for support under the Consolidated Grant ST/S000623/1, as well as the European Research Council (ERC) for support under the European Union Horizon 2020 research and innovation programme (grant 834203).~M.L.P.~is supported by European Space Agency (ESA) Research Fellowships.~J.H.M acknowledges a Herchel Smith Fellowship while at Cambridge and support from the Royal Society at Oxford.~M.C.D.M.~is supported by the European Research Council under Grant No.~742104 and by the Swedish Research Council (VR) under grants 2018-03641 and 2019-02337.~This work was performed using resources provided by the Cambridge Service for Data Driven Discovery (CSD3) operated by the University of Cambridge Research Computing Service (\url{www.csd3.cam.ac.uk}), provided by Dell EMC and Intel using Tier-2 funding from the Engineering and Physical Sciences Research Council (capital grant EP/P020259/1), and DiRAC funding from the Science and Technology Facilities Council (\url{www.dirac.ac.uk}).~This work was performed using resources provided by the Cambridge Service for Data Driven Discovery (CSD3) operated by the University of Cambridge Research Computing Service (www.csd3.cam.ac.uk), provided by Dell EMC and Intel using Tier-2 funding from the Engineering and Physical Sciences Research Council (capital grant EP/P020259/1), and DiRAC funding from the Science and Technology Facilities Council (www.dirac.ac.uk).~We gratefully acknowledge the use of the following software packages: astropy \citep{astropy2013,astropy2018}, matplotlib \citep{matplotlib_07}, pandas \citep{mckinney-proc-scipy-2010, reback2020pandas_latestV_zenodo}, OpenMPI \citep{gabriel04_open_mpi}, and \textsc{xspec} and \textsc{PyXspec} \citep{Xspec_LatestRef_1996Arnaud}.~As introduced in Sec.~\ref{section:s4_alpGrid}, we have employed the public-source code \alpro\ v1.0 \citep{alpro_james_2022_zenodo}.


\bibliography{myReferences.bib}



\appendix
\section{Theory of photon-ALP mixing}
\label{section:a1_theory_photonALPmixing}
At a fundamental level, the interaction between ALPs and electromagnetism is specified by Eq.~\ref{equation:eq1_interactionLangrangian_alps}.~Here, we present the relevant equations derived from Eq.~\ref{equation:eq1_interactionLangrangian_alps} for a simplified external field geometry.~This is for illustration purposes and context only.~We refer to Secs.~2 and 3 of \citet[][]{2022_matthews_bfieldModels} for a description of the calculations used to generate the photon-ALP mixing grid employed throughout our analysis.~

We proceed to consider a scenario where the electromagnetic wave is traversing an ionised and magnetised medium such that its self-induced magnetic field can be ignored relative to the latter.~We refer to \citet[][]{Galanti_2022_overview_Universe_BfieldStrong} for an overview of the relevant equations inferred when the self-induced electric and magnetic fields are not negligible.

We consider an initially unpolarized beam propagating through an external homogeneous field perpendicular to the direction of propagation, $\textbf{B}_0$ (of strength $B_0$).~Most generally, however, the magnetic field geometry of a given external medium will be complex, for which the photon-ALP mixing problem (derived from Eq.~\ref{equation:eq1_interactionLangrangian_alps}) must be solved numerically.

In the field geometry considered, a wave-like equation can be derived for $\mathcal{L}_\mathrm{a\gamma}$ which, at energies $E >> m_\mathrm{a}$, for {$m_\mathrm{a} \leq 10^{-10} \ \mathrm{eV}$, simplifies to a Schrodinger-type equation that can be solved through the eigenvalue problem.~The solution to such problem can be quantified via a non-unity ``survival'' probability for the photon beam as it interconverts into ALPs of mass $m_\mathrm{a}$ and coupling $g_\mathrm{a\gamma}$ after having travelled distance $L$ along the magnetised medium.~Following Sec.~4.1 of \citet[][]{Marsh17}, $P_\mathrm{\gamma \rightarrow \gamma}$ is given by: \begin{equation}
  \label{equation:eq2_prob_inter-conversion}
  \centering 
    P_\mathrm{\gamma \rightarrow \gamma} = 1 - \frac{1}{4} \frac{1}{1 + \theta^{-2}} ~ \mathrm{sin}^{2} \bigg(L ~ \Delta_\mathrm{eff} \sqrt{1 + \theta^{2}} \ \bigg),
\end{equation} where $\theta$ and $\Delta_\mathrm{eff}$ are defined via: \begin{equation}
  \label{equation:eq3_thetaFactor}
  \centering 
  \theta = 4\pi ~ \bigg(\frac{B_0 ~ g_\mathrm{a\gamma} ~ E}{m_\mathrm{a}^2 - \omega_\mathrm{pl}^{2}}\bigg) 
\end{equation} and 
\begin{equation}
  \label{equation:eq4_effectiveLength_alp}
  \centering 
  \Delta_\mathrm{eff} = \frac{\omega_\mathrm{pl}^{2} - m_\mathrm{a}^{2}}{8\pi E} 
\end{equation} where $\omega_\mathrm{pl}$ is the plasma frequency of the ionised medium and $m_\mathrm{eff}^{2} = m_\mathrm{a}^{2} - \omega_\mathrm{pl}^{2}$ is the effective photon mass.~

For effectively massless ALPs ($m_\mathrm{a}<<\omega_\mathrm{pl}$, i.e.$~m_\mathrm{eff}^2 \rightarrow -\omega_\mathrm{pl}^2$), we now proceed to consider the regime where Eqs.~\ref{equation:eq3_thetaFactor} and \ref{equation:eq4_effectiveLength_alp} satisfy: $\theta >> 1$ and $\Delta_\mathrm{eff} << 1$, respectively.~In a suitable environment (i.e.~that provided by rich cool-core clusters and for a suitable location of ALP parameter space), the survival function will be:
\begin{equation}
    \centering
    \label{equation:app_mixing_clusters}
    P_\mathrm{\gamma \rightarrow \gamma } \sim 1 - 10^{-3} ~{\bigg(\frac{B_0}{1 \ \mu \mathrm{G}}\bigg)}^{2}{\bigg(\frac{g_\mathrm{a\gamma}}{10^{-10} \ {\mathrm{GeV}}^{-1}}\bigg)}^{2} ~ {\bigg(\frac{L}{1 \ \mathrm{kpc}}\bigg)}^{2} ~ g(E),
\end{equation}
where we have adopted a similar notation to \citet[][]{Conlon_2017_ProjectedBoundsOnAthena} and where $g(E) \in [0,1]$ is an oscillatory function with energy of order unity.~We highlight that the product $(B_0 ~ L)$ in Eq.~\ref{equation:app_mixing_clusters} can attain large values in rich cool-core clusters compared to other astrophysical systems, which underlies why rich clusters should be efficient photon-ALP interconverters.
\section{Effects of non-optimal calibration on \textit{Athena}/X-IFU revisited}
\label{section:appendix_illustration_calibration}
The left panel of Fig.~\ref{figure:f4_residuals_differentARFs}~shows the residuals resulting from fitting the simulated \textit{Athena}/X-IFU spectrum of NGC1275 with an astrophysics-only spectral model (see Tab.~\ref{table:t2_astroModel_specifications_baseline}), i.e.~without ALPs.~The 
input spectrum is then interpreted under an ``optimal'' calibration (simulating and fitting with the same ARF) and an example ``non-optimal'' calibration (simulating and fitting with different ARFs) scenarios (left and central panels, respectively).~Clearly, the miscalibrated spectrum seems to retain most of the curvature present in the residuals of the optimal case with the exception of acquiring additional curvature at energies $< 2 \ \mathrm{keV}$.~This is likely due to a miscalibration of instrumental effects at such energies.~We highlight that an even more conservative approach to assessing the effects of detector calibration would involve convolving the miscalibrated responses for the mirror and microcalorimeters, but is, however, beyond the scope of this paper \citep[see discussion in][]{Barret_and_cappi_2019_bhspin_degen}.
\begin{figure}
 \center
 \includegraphics[width=0.45\textwidth]{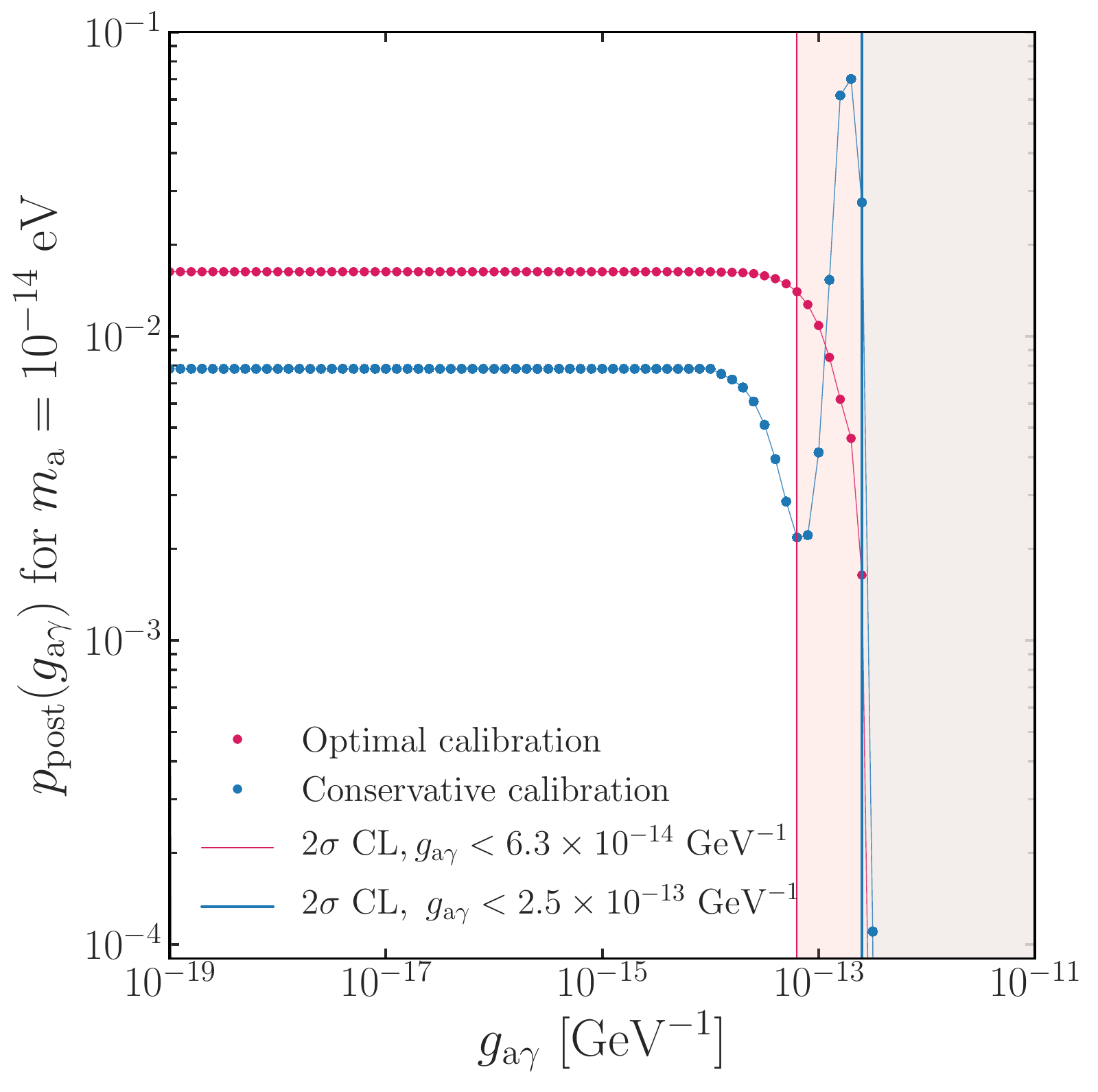}
 \caption{Normalised posterior distribution on $g_\mathrm{a\gamma}$ for ALPs of $m_\mathrm{a} = 10^{-14} \ \mathrm{eV}$ (see Eq.~\ref{eq:eq6_normalisation_condition}), inferred when fitting our simulated \textit{Athena}/X-IFU spectrum of NGC1275 in the ``optimal'' and a representative ``conservative'' detector response scenarioes (Sec.~\ref{section:s6.1_importance_arf_illustrated} for discussion).~The posterior is inferred by marginalising over 216 magnetic field realisations.~The shaded regions delimit regions of parameter space that are excluded at $95\% \ (\mathrm{i.e.} \ 2\sigma) \ \mathrm{confidence}$, respectively.}
 \label{figure:f6_posteriors}
\end{figure} The right panel of Fig.~\ref{figure:f4_residuals_differentARFs} shows the residuals resulting from fitting a multiplicative ALP model in our photon-ALP mixing grid (Sec.~\ref{section:s4_alpGrid}) to the simulated \textit{Athena}/X-IFU spectrum of NGC1275 under the optimal calibration scenario.~For the specific turbulent field realisation and value of $g_\mathrm{a\gamma}$ chosen, the ALP-induced features result in well-defined peaks in the residuals whose amplitude increases with energy.~Interestingly, the $C$-stat of such ALP fit will still more accurately describe the simulated spectrum of NGC1275 compared to a miscalibrated non-ALP containing model.~This is illustrated by comparing the $C$-stats of the central and right panels to that in the left panel of Fig.~\ref{figure:f4_residuals_differentARFs}.

For further comparison, we fit the miscalibrated {Athena}/X-IFU spectrum (Sec.~\ref{section:s3_simulated_spectra_ngc1275_nominal}) across our grid of photon-ALP mixing models \{\texttt{ALP(i,$g_\mathrm{a\gamma}$)}\}, following the procedure described in Sec.~\ref{section:s5_nominal_nextGen}.~The marginalised posterior (over 216 field models) is compared to that for the optimal calibration case in Fig.~\ref{figure:f6_posteriors}.~Importantly, we see that the former is now dominated by large values of the photon-ALP coupling $g_\mathrm{a\gamma}$ (i.e.~above the $95\%$ exclusion level in the optimal calibration case) suited to describe the energy-dependent features induced by instrumental detector miscalibration.~The analysis of such a miscalibrated spectrum would translate into a weakening of the 95\%~$\mathrm{CL}$ on $g_\mathrm{a\gamma}$ by a factor of 4, compared to the optimal calibration scenario.~In Sec.~\ref{subsection:s6.2_marginalising_over_differentARFs}, we thereby proceed to assess the potential effects of detector miscalibration on the resulting posteriors on ALPs by considering a library of conservative ARFs.

\section{Marginalising over response files}
\label{section:a2_bayesianApproach_calibration_marginalisation}

In Sec.~\ref{section:s5_nominal_nextGen}, we present the optimal bounds on ALPs from \textit{Athena}/X-IFU and \textit{AXIS} by using the Bayesian framework presented in Sec.~4 of \citet[][]{sisk21_alps}.~We hereby proceed to compute \textit{Athena}'s bounds on massless ALPs by marginalising over a set of $216$ area response files (ARFs) and $216$ realisations of the turbulent field within Perseus.~Our results are presented in Sec.~\ref{subsection:s6.2_marginalising_over_differentARFs}.

Equipped with the library of photon-ALP mixing curves introduced in Sec.~\ref{section:s4_alpGrid}, we proceed as follows.~We iteratively fit the optimal \textit{Athena}/X-IFU spectrum of NGC1275 (Fig.~\ref{figure:f2_simulated_xifu_axis_chandra}) with the spectral model \texttt{tbabs*(po + bvvapec)*ALP($m_\mathrm{a}$, $g_\mathrm{a\gamma}$, i} across a library of conservative detector responses, $\{\texttt{f}\}$.~Here,~\texttt{ALP($m_\mathrm{a}$,$g_\mathrm{a\gamma}$,i)}~quantifies the energy-dependent survival function of quanta mixing with ALPs of parameters ($m_\mathrm{a},g_\mathrm{a\gamma}$) for a given field realisation \texttt{i}.~We refer to the goodness-of-fit statistic ($C$-stat) of each fit as $C(m_\mathrm{a},g_\mathrm{a\gamma},\texttt{i},\texttt{f})$.~The free parameters of each fit are outlined in Tab.~\ref{table:t2_astroModel_specifications_baseline}.~Each multiplicative table \texttt{ALP} model has a redshift parameter attributed to it, frozen to that of the galaxy \citep[$z = 0.01278$,][]{2018_hitomi_citation_for_z}.

We now introduce the Bayesian framework that permits converting $C(m_\mathrm{a},g_\mathrm{a\gamma}, \texttt{i},\texttt{f})$ into a normalised posterior $p_\mathrm{post}(m_\mathrm{a},g_\mathrm{a\gamma})$.~The latter can then be used to find confidence levels on ALP parameter space.

Ultimately, we seek to find a posterior distribution on ALP parameters $p_\mathrm{post}(m_\mathrm{a}, g_\mathrm{a\gamma})$~normalised according to the condition:\begin{equation}
  \centering
  \label{eq:eq6_normalisation_condition}
  \sum_{\mathrm{log}(m_\mathrm{a}/\mathrm{eV})} \ \sum_{\mathrm{log}(g_\mathrm{a\gamma}/\mathrm{GeV}^{-1})} {p}_\mathrm{post}(m_\mathrm{a}, g_\mathrm{a\gamma}) = 1,
\end{equation}where we assume flat priors on the ALP parameters.~The summations involved comprise the parameter space: $\mathrm{log}(m_\mathrm{a}/\mathrm{eV}) \in [-30.0, -12.0) \ \bigcup \ \mathrm{log}(g_\mathrm{a\gamma}/{\mathrm{GeV}}^{-1}) \in [-19.0, -11.0]$, where the choice of lower limits in $m_\mathrm{a}$ and $g_\mathrm{a\gamma}$ is justified in Sec.~\ref{section:s5_nominal_nextGen}.

Operationally, $p_\mathrm{post}(m_\mathrm{a}, g_\mathrm{a\gamma})$ is found by marginalising over all magnetic field models $\texttt{i} \in [1,216]$ and detector responses $\texttt{f} \in [0, 215]$, where \texttt{f0} is the optimal \textit{Athena}/X-IFU ARF.~We broadly follow the description in A2 of \citet[][]{Marsh17}, where we marginalise over the added degree-of-freedom encapsulated by the set of ARFs (\texttt{f}).~For a given ARF \texttt{f}, the unnormalised posterior on a given set of ALP parameters $(m_\mathrm{a}, g_\mathrm{a\gamma})$ is given by: \begin{equation}
  \centering
  \label{eq:eq7_posterior_singlek_marg_overi}
  {p}_\mathrm{post}(m_\mathrm{a}, g_\mathrm{a\gamma}, \texttt{f}) \propto \sum_{\texttt{i} = 1}^{216} \mathrm{exp} \biggl\{ \frac{ - \big ( C(m_\mathrm{a},g_\mathrm{a\gamma},\texttt{i},\texttt{f}) - C_\mathrm{fid}(\texttt{f}) \big)}{2} \biggl\},
\end{equation} where the summation over $\texttt{i}$~marginalises over magnetic field configurations, on which we assume flat priors \citep[as previously done in][]{sisk21_alps, 2022_matthews_bfieldModels}.~Here, $C_\mathrm{fid}(\texttt{f})$ is the ``fiducial'' best-fit $C$-stat of the simulated X-IFU spectrum read on the $\texttt{f}$-th ARF when fitted with the \texttt{tbabs*(po+bvvapec)} model (see Tab.~\ref{table:t2_astroModel_specifications_baseline}).

Clearly, for a given ($\texttt{f}$-th) ARF, the posterior distribution on ALPs will be dominated by the ALP models suited to describing the underlying residuals in the spectrum of NGC1275 present once its calibration-induced features have been corrected.~Similar to the case of the magnetic field realisation, the dependence on detector response can be eliminated through: 
\begin{equation}
  \centering
  \label{eq:eq8_posterior_margOverARFs}
  p_\mathrm{post}(m_\mathrm{a}, g_\mathrm{a\gamma}) \equiv \mathcal{L}(m_\mathrm{a}, g_\mathrm{a\gamma}) = \sum_{\texttt{f} = 0}^{215}{p}_\mathrm{post}(m_\mathrm{a},g_\mathrm{a\gamma},\texttt{f}) 
\end{equation} where, again, we assume flat priors on the ARF nuisance parameter.~Operationally, we find the constant of proportionality in Eq.~\ref{eq:eq7_posterior_singlek_marg_overi} through the normalisation condition of $p_\mathrm{post}(m_\mathrm{a}, g_\mathrm{a\gamma})$, i.e.~Eq.~\ref{eq:eq6_normalisation_condition}.~In the equation above, we have explicitly stated the equivalence of the normalised posterior and the marginalised likelihood $\mathcal{L}(m_\mathrm{a}, g_\mathrm{a\gamma})$ over ARFs and field configurations.

Finally, we note the ALP constraints inferred in our work are insensitive to the choice of binning scheme chosen throughout the data processing discussed in Secs.~\ref{section:s5_nominal_nextGen} and \ref{section:s6_bounds_differentCalibration}.
\label{lastpage}
\end{document}